\documentclass[twoside,twocolumn,9pt]{article}
\usepackage{extsizes}
\usepackage[super,sort&compress,comma]{natbib} 
\usepackage[version=3]{mhchem}
\usepackage[left=1.5cm, right=1.5cm, top=1.785cm, bottom=2.0cm]{geometry}
\usepackage{balance}
\usepackage{mathptmx}
\usepackage{sectsty}
\usepackage{graphicx} 
\usepackage{lastpage}
\usepackage[format=plain,justification=justified,singlelinecheck=false,font={stretch=1.125,small,sf},labelfont=bf,labelsep=space]{caption}
\usepackage{float}
\usepackage{fancyhdr}
\usepackage{fnpos}
\usepackage[english]{babel}
\addto{\captionsenglish}{%
  
}
\usepackage{array}
\usepackage{droidsans}
\usepackage{charter}
\usepackage[T1]{fontenc}
\usepackage[usenames,dvipsnames]{xcolor}
\usepackage{setspace}
\usepackage[compact]{titlesec}
\usepackage{hyperref}
\usepackage{chemfig}
\usepackage{xcolor}

\usepackage{epstopdf}
\usepackage{subcaption}

\definecolor{cream}{RGB}{222,217,201}

\begin{document}

\pagestyle{fancy}
\thispagestyle{plain}
\fancypagestyle{plain}{
\renewcommand{\headrulewidth}{0pt}
}

\makeFNbottom
\makeatletter
\renewcommand\LARGE{\@setfontsize\LARGE{15pt}{17}}
\renewcommand\Large{\@setfontsize\Large{12pt}{14}}
\renewcommand\large{\@setfontsize\large{10pt}{12}}
\renewcommand\footnotesize{\@setfontsize\footnotesize{7pt}{10}}
\makeatother

\renewcommand{\thefootnote}{\fnsymbol{footnote}}
\renewcommand\footnoterule{\vspace*{1pt}%
\color{cream}\hrule width 3.5in height 0.4pt \color{black}\vspace*{5pt}} 
\setcounter{secnumdepth}{5}

\makeatletter 
\renewcommand\@biblabel[1]{#1}            
\renewcommand\@makefntext[1]%
{\noindent\makebox[0pt][r]{\@thefnmark\,}#1}
\makeatother 
\renewcommand{\figurename}{\small{Fig.}~}
\sectionfont{\sffamily\Large}
\subsectionfont{\normalsize}
\subsubsectionfont{\bf}
\setstretch{1.125} 
\setlength{\skip\footins}{0.8cm}
\setlength{\footnotesep}{0.25cm}
\setlength{\jot}{10pt}
\titlespacing*{\section}{0pt}{4pt}{4pt}
\titlespacing*{\subsection}{0pt}{15pt}{1pt}

\fancyfoot{}
\fancyfoot[LO,RE]{\vspace{-7.1pt}\includegraphics[height=9pt]{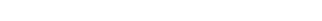}}
\fancyfoot[CO]{\vspace{-7.1pt}\hspace{11.9cm}\includegraphics{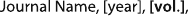}}
\fancyfoot[CE]{\vspace{-7.2pt}\hspace{-13.2cm}\includegraphics{head_foot/RF}}
\fancyfoot[RO]{\footnotesize{\sffamily{1--\pageref{LastPage} ~\textbar  \hspace{2pt}\thepage}}}
\fancyfoot[LE]{\footnotesize{\sffamily{\thepage~\textbar\hspace{4.65cm} 1--\pageref{LastPage}}}}
\fancyhead{}
\renewcommand{\headrulewidth}{0pt} 
\renewcommand{\footrulewidth}{0pt}
\setlength{\arrayrulewidth}{1pt}
\setlength{\columnsep}{6.5mm}
\setlength\bibsep{1pt}

\makeatletter 
\newlength{\figrulesep} 
\setlength{\figrulesep}{0.5\textfloatsep} 

\newcommand{\topfigrule}{\vspace*{-1pt}%
\noindent{\color{cream}\rule[-\figrulesep]{\columnwidth}{1.5pt}} }

\newcommand{\botfigrule}{\vspace*{-2pt}%
\noindent{\color{cream}\rule[\figrulesep]{\columnwidth}{1.5pt}} }

\newcommand{\dblfigrule}{\vspace*{-1pt}%
\noindent{\color{cream}\rule[-\figrulesep]{\textwidth}{1.5pt}} }

\makeatother

\twocolumn[
  \begin{@twocolumnfalse}
{\includegraphics[height=30pt]{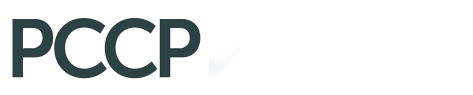}\hfill\raisebox{0pt}[0pt][0pt]{\includegraphics[height=55pt]{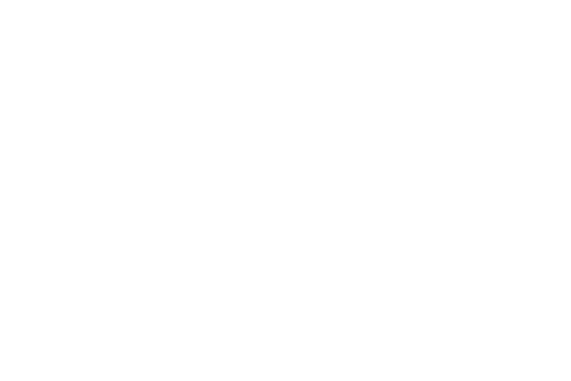}}\\[1ex]
\includegraphics[width=18.5cm]{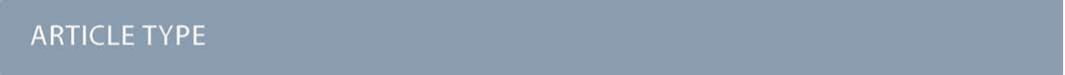}}\par
\vspace{1em}
\sffamily
\begin{tabular}{m{4.5cm} p{13.5cm} }

\includegraphics{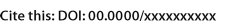} & \noindent\LARGE{\textbf{Photofragmentation of Corannulene (C\textsubscript{20}H\textsubscript{10}) and Sumanene (C\textsubscript{21}H\textsubscript{12}) cations in gas phase and its Astrophysical implications $^\dag$}} \\
\vspace{0.3cm} & \vspace{0.3cm} \\

 & \noindent\large{Pavithraa Sundararajan,$^{\ast}$\textit{$^{,a}$}\textit{$^{,b}$} Alessandra Candian,\textit{$^{c}$}Jerry Kamer, \textit{$^{a,}$}\textit{$^{b}$} Harold Linnartz,\textit{$^{a,}$}\textit{$^{b}$}and Alexander G.G.M. Tielens\textit{$^{b,}$}\textit{$^{d}$}} \\

\includegraphics{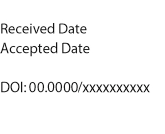} & \noindent\normalsize{The aromatic infrared Bands (AIBs) dominate the mid-infrared spectra of many galactic and extragalactic sources. These AIBs are generally attributed to fluorescent emission from aromatic molecules. Unified efforts from experimentalists and theoreticians to assign these AIB features have recently gotten additional impetus with the launch of the James Webb Space Telescope (JWST) as the Mid-InfraRed Instrument (MIRI) delivers mid-IR spectrum with greatly increased sensitivity and spectral resolution. PAHs in space can exist in either neutral or ionic forms, absorb UV photons and undergo fragmentation, becoming a rich source of small hydrocarbons. This top-down mechanism of larger PAHs fragmenting into smaller species is of utmost importance in photo-dissociation regions (PDR) in space. In this work, we experimentally and theoretically investigate the photo-fragmentation pathways of two astronomically significant PAH cations – corannulene (C\textsubscript{20}H\textsubscript{10}) and sumanene (C\textsubscript{21}H\textsubscript{12}) that are structural motifs of fullerene C\textsubscript{60}, to understand their sequential fragmentation pathways. The photo-fragmentation experiments exhibit channels that are much different from planar PAHs. The breakdown of carbon skeleton is found to have different pathways for C\textsubscript{20}H\textsubscript{10} and C\textsubscript{21}H\textsubscript{12} because of the number and positioning of pentagon rings; yet the most abundant low mass cations produced by these two species are found to be similar. The low mass cations showcased in this work could be of interest for their astronomical detections. For completeness, the qualitative photo fragmentation behaviour of the dicationic corannulene and sumanene have also been experimented, but the potential energy surface of these dications are beyond the scope of this paper.} \\

\end{tabular}

 \end{@twocolumnfalse} \vspace{0.6cm}

  ]

\renewcommand*\rmdefault{bch}\normalfont\upshape
\rmfamily
\section*{}
\vspace{-1cm}


\footnotetext{\textit{$^{a}$~Laboratory for Astrophysics, Leiden University, PO Box 9513, NL-2300, RA Leiden, the Netherlands.; E-mail: sundararajan@strw.leidenuniv.nl}}
\footnotetext{\textit{$^{b}$~Leiden Observatory, Leiden University, 2300 RA Leiden, The Netherlands.}}
\footnotetext{\textit{$^{c}$~Anton Pannekoek Institute, University of Amsterdam, Science Park 904, 1098XH Amsterdam, The Netherlands.}}
\footnotetext{\textit{$^{d}$~Astronomy Department, University of Maryland, College Park, MD 20742, USA.}}

\footnotetext{\dag~Electronic Supplementary Information (ESI) available: See DOI: 10.1039/cXCP00000x/}


\section{Introduction}
The AIBs are discrete emission features observed in the interstellar medium (ISM), circumstellar regions, galactic and extragalactic sources.\cite{candian2018aromatic} The AIBs are very characteristic for the \chemfig{C - C} and \chemfig{C - H} vibrational modes of polycyclic aromatic hydrocarbons (PAHs) with prominent features cantered at 3.3, 6.2, 7.7, 8.6, 11.3, and 12.7~$\mu$m region. These bands are attributed to IR fluorescence of UV pumped PAHs species that after internal conversion and intramolecular vibrational energy distribution relax through vibrational emission. PAHs can occur in the ISM either in the neutral, cationic, protonated, hydrogenated, dehydrogenated or substituted (with one or more carbon atoms replaced with atoms like nitrogen, oxygen, etc.) forms.\cite{peeters2021spectroscopic}  The wavelength and relative intensity of these astronomical AIBs are found to vary with physical environment.\cite{joblin2011shape}  In spite of numerous theoretical and experimental investigations for several decades to deduce the emitters of the AIB’s, the exact form of PAHs or the group of PAHs responsible for the emission continues to puzzle astronomy. Recent secure identifications of small PAHs in dark clouds through their rotational spectra has confirmed the presence of this important class of molecules in space.\cite{mcguire2021detection}

PAHs play an important role in the ionization and energy balance of interstellar gas.\cite{tielens1995interstellar}  In addition, observations expose that larger PAHs (with 20 to 100 carbon atoms) harbours 10–15 \% of the elemental carbon in the ISM. Besides PAHs, the buckminsterfullerene, \ce{C60}, has also been identified in the ISM through its infrared bands using the Spitzer Space Telescope\cite{cami2010detection}, and this was confirmed through its electronic transitions in the far-red.\cite{campbell2015laboratory,campbell2020spectroscopy}  It is found that, close to stars, the abundance of \ce{C60} increases rapidly while the abundance of PAHs decreases.\cite{tielens2005physics,berne2012formation} This has been attributed to photochemical fragmentation and isomerization processes under the influence of the strong stellar ultraviolet (UV) radiation field but the details of these processes are not yet understood. A detailed study of the bright reflection nebula, NGC 7023,  has given evidence that the evolution of the profile of the mid-IR bands is related to the chemical evolution under the effect of UV photons.\cite{joblin2011shape}

PAHs are assumed to be formed in the envelopes of evolved stars and then to be injected in the ISM.\cite{frenklach1989formation,cherchneff1992polycyclic} In the past, astronomical models generally postulated that the abundance of specific PAHs in the ISM was mainly controlled by their thermodynamic properties and followed the pattern of stabilomers originally identified by Stein and Fahr (1985).\cite{stein1985high}  However, under the severe conditions of space where there are strong radiation fields, photochemical pathways may control the abundances of interstellar PAHs. In this scenario, it is considered that stars inject a rich variety of PAHs – formed through processes akin to those in sooting flames ,  – into space where they are quickly weeded down to a small set of extremely stable species by the strong radiation field. PAHs in space typically absorb ~108 UV photons over their lifetime (~100 Myr) and undergo destruction through photo-fragmentation process.\cite{tielens2005physics} The composition of the interstellar PAH family is greatly influenced by the fragmentation process initiated by UV photon absorption as highly excited PAHs can lose H atoms, CH, and \ce{C2}/\ce{C2H2} groups rather than relax through IR emission. This competition between fragmentation and radiative relaxation is largely controlled by the size of PAH.\cite{berne2012formation,berne2015top,croiset2016mapping,west2019large,zhen2014laboratory} 

Recent laboratory experiments demonstrate that upon photo-excitation the fragmentation of PAHs offers a new chemical paradigm of complex, interstellar molecule formation; Instead of bottom-up, where reacting smaller species form larger ones,  a top-down mechanism, where fragmentation products of larger precursors occur.\cite{zhen2014laboratory,zhen2014quadrupole} The observed anti-correlation between PAHs and \ce{C60} abundances in astronomical environments strongly suggests that one species forms from the other and this photochemical relationship carries the potential to give us direct clues on the processes that are dominant in the evolution of interstellar PAHs. The fullerenes contain fused hexagon and pentagon rings, and pentagon formation may provide the key.\cite{ota2017reproduction} In an intense UV radiation environment, this photochemical breakdown may also contribute to the diversity of small hydrocarbon radicals commonly observed in these regions.\cite{pety2005pahs} Understanding the dissociation of PAHs through photochemical excitation requires dedicated laboratory studies to identify the detailed photo-fragmentation of PAHs in space.

Neutral and cationic \ce{C60} – identified through their unique IR and visible spectrum\cite{cami2010detection,campbell2015laboratory} – are the largest molecular species identified in space. The presence of pentagons in a planar graphene-flake leads to bending of the molecular structure enabling the closure upon itself and formation of a 3D structure.\cite{chen2020molecular,kroto1987jr} Pentagon formation is thus a key step in the photochemical transformation of PAHs into fullerenes. Spectroscopic evidence for pentagon formation upon PAH photo-dissociation has been reported in the laboratory for small PAH species, containing up to three aromatic rings.\cite{bouwman2016spectroscopic,de2017facile,banhatti2022formation} These experiments suggest that PAHs with only hexagonal rings could be the parent molecules (Parent-PAHs) for the formation of PAHs with pentagonal ring(s) in adequate conditions.  In addition, some PAHs with pentagons may be more photochemically stable than PAHs of similar size\cite{ekern1998photodissociation} and therefore represent likely candidates for the interstellar molecular inventory. However, the photochemical evolution of PAHs with pentagons is largely unexplored and the generality of this conclusion is unknown. Given the presence of \ce{C60} in space, the likely astrochemical connection of \ce{C60} and PAHs in space, and the potential photochemical stability of pentagon containing PAHs, a focused study to improve the understanding of such species is portrayed in this work.

\begin{figure}[h]
\centering
  \includegraphics[height=5.9cm]{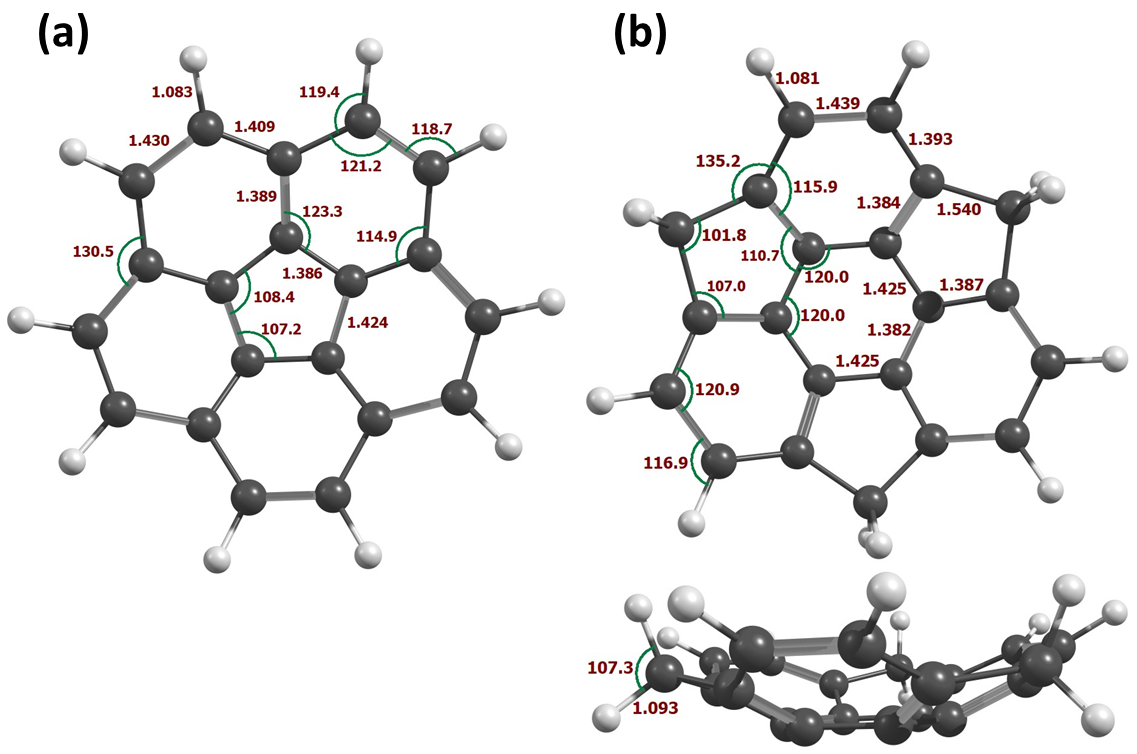}
  \caption{Geometry of cationic (a) Corannulene, \ce{C20H10+} and (b) Sumanene, \ce{C21H12+} optimized using M062x/6-311++G(3df,2pd) method.}
  \label{Geometry}
\end{figure}

PAHs with pentagons chosen for this work are: (a) corannulene (\ce{C20H10}), and (b) sumanene (\ce{C21H12}) as shown in Figure \ref{Geometry}. The coordinates of the optimized geometry have been provided in Figure S1 and S2 in the supporting information. These PAHs have been selected because they represent "pentagon-variations" on the structure of the well-studied compact PAH, coronene. Moreover, corannulene and sumanene are molecular building blocks/fragments of \ce{C60} as the pentagon is connected to five hexagons in corannulene and the pentagons in sumanene are separated from each other through (partial) rings of hexagons.\cite{narahariasastry1993synthetic} When corannulene has ten H atoms which are all sp$^2$ hybridized, sumanene has six H toms that are sp$^2$ and six more H atoms that are sp$^3$ hybridized (figure \ref{Geometry} bottom panel). This makes the chemistry of these two molecules much different from each other. The photochemistry of the Parent-PAH, coronene as well as the fullerenes \ce{C60} and \ce{C70}, have been studied in detail in the gas phase, however, the buckybowls (\ce{C20H10} and \ce{C21H12}) are least explored.\cite{oomens2001gas,zhen2016photo}

Several spectroscopic studies have been performed on corannulene to look for its presence in the interstellar medium.\cite{gatchell2019spectroscopy,Rouille2008,perez2017corannulene} In particular, the diffuse interstellar bands at 6614 and 6196 Å have been proposed to arise from the electronic states of corannulene.\cite{bernstein2018spectral} Moreover, the matrix-isolated IR spectra of hub-protonated corannulene (\textit{hub}-\ce{H+C21H10+}) showed close resemblance with the astronomical UIR bands,\cite{sundararajan2018infrared} but gas-phase spectra are required for positive astronomical detection. Because of its non-planar structure,\cite{weber2022infrared} corannulene possesses a large dipole moment (2.0 Debye) and its microwave spectrum has been recorded.\cite{lovas2005interstellar} A detailed model suggest that this species could be present in the Red Rectangle nebula, yet the hunt for this molecule still continues.\cite{pilleri2009search} 
The spectroscopy of sumanene is not so well constrained. The molecule has a slightly larger bowl depth (1.11 Å) than corannulene (0.87 Å)  and it is closer to the curvature of the \ce{C60} structure.\cite{narahariasastry1993synthetic,lawton1971synthesis} The presence of three sp\textsuperscript{3}-hybridized benzylic sites is a unique feature in sumanene, which makes its chemistry much different from that of corannulene.\cite{amaya2015chemistry} Sumanene is expected to be an important precursor for the formation of \ce{C60} via bottom-up mechanism.
 
To date, a number of laboratory experiments have been dedicated to study the photochemistry of smaller PAHs (C<20).\cite{pety2005pahs,lifshitz1997energetics,garkusha2011electronic,joblin1994ir} Experimental studies of PAHs relevant to astrochemistry have widely used a time-of-flight, photo-ionization mass spectrometry (PIMS) method, focusing on reaching internal energy at which the dissociation rate is $\sim 10^{4}$ s$^{-1}$.\cite{jochims1999structure}
 
In these experiments, loss channel(s) mainly involve sequential hydrogen loss followed by the opening up of the \ce{C2H2} loss channel at slightly higher internal energies. However, different molecular dynamical timescales are showcased in the ISM. A competition takes place between fragmentation of PAHs, isomerization, and relaxation through IR emission that occurs on time scales of the order of $\sim 0.01-1$ s$^{-1}$, depending on the size of the species. This work studies in detail the photofragmentation pattern of the buckybowls, the influence of the presence of pentagon rings on the photo-dissociation pattern, the isomerization process involved in formation of smaller fragment cations with the aid of the mass spectroscopic data obtained in the ‘instrument for Photo-dynamics of PAHs' (i-PoP) system. 

\section{Experiment}

The experiments presented in this paper were performed on the ‘instrument for Photo-dynamics of PAHs’ (i-PoP), housed at the Laboratory for Astrophysics (LfA) at Leiden observatory. i-PoP has been described in detail in a previous paper so only a brief description is given here.\cite{zhen2014quadrupole} The setup consists of two differentially pumped chambers; a source chamber that has a commercially available quadrupole ion trap (QIT, Jordan C-1251), and a detection chamber which encompasses a reflectron time-of-flight spectrometer (Jordan D-850). The molecules used in this work are commercially available from Tokyo Chemical Industry; Corannulene (\ce{C20H10}, >97.0\% pure), and Sumanene (\ce{C21H12}, 99.0\% pure). The samples were evaporated in the QIT chamber using a Heat Wave Labs built oven held at around 40 °C and 70 °C for corannulene and sumanene respectively. The sample vapours were ionized by an electron impact ionization at ~70 eV with an electron gun (EGUN, Jordan C-950) which is integrated to the QIT chamber and the resulting cations after ionization were then guided into the ion trap through an electrostatic ion gate.

A 1600 V top-top RF signal were used for these experiments on the ring electrode at an operating frequency of 1.25 MHz in the ion trap to hold ions. This enables the trap to retain masses from approx. 98 amu up to several hundred amu. In an attempt to probe a much lower mass range to look for smaller cationic fragments and RF of 1200 V was also used in some experiments which pushed the lower observable limit to ~73 amu. In addition, the experiments with dications uses a RF of 1000 V to observe till ~61 amu.  Helium buffer gas was let into the center of the ion trap up to a static pressure of 1 – 2 x 10\textsuperscript{-6} mbar in the QIT chamber. The PAH cations were confined to the center of the ion trap through collisions with the He buffer gas and persisted there until the cations were directed from the ion trap into the time-of-flight (TOF) detection chamber.\cite{sassin2008photodissociation,gulyuz2011hybrid} The ions in the ion trap were irradiated with a nanosecond pulsed Quanta-Ray Nd:YAG laser (DCR2A-3235) pumping a dye laser (LIOP-TEC, Quasar2-VN) which was set to deliver a particular wavelength from 610 – 630 nm photons chosen depending on the molecule to be studied. The laser was horizontally guided through the ion trap and was operated at 10 Hz to irradiate the trapped ions.

The reasoning for the choice of 610 – 630 nm (red light) laser radiation was to scan over the potential energy surface of corannulene and sumanene cations which was the main aim of this experiment. The choice of red light from dye laser radiation allows to minimize multiple ionization in competition with fragmentation that is known to occur with VUV photons.\cite{zhen2016vuv,wenzel2020astrochemical} For, instance, using a green light for photolysis would give rise to the same set of product cations. But, there are higher chances of the cations undergoing multiple ionization, making the analysis complicated. Moreover, the breakdown analysis to deduce the steps of fragmentation would not be possible as several fragment cations are produced in higher intensities at the very beginning of the laser pulses. Hence, using longer wavelength photons in a multi-photon process allowed to obtain more information about the fragmentation pattern of the molecules in a gentler manner to explicate additional nuances in the fragmentation patterns.

The data acquisition cycle is achieved through timing sequences controlled by a high-precision delay generator (SRS DG535), which is triggered by the Q-switch timing of the Nd:YAG laser to ensure synchronization at the start of each measurement cycle. Each operation cycle involves capturing ions and filling of the ion trap, mass isolation of the parent ion, irradiation of the ion cloud with laser, and extraction of the ions into the TOF tube. The scan cycle begins with an empty ion trap and was initiated with the opening of the ion gate which is done by normal gating procedure of applying a DC voltage to a metal lens with a circular slit in the middle to inlet the ions, and the ion trap fills for a duration of 2.7 s. To isolate the masses of the parent PAH ions, a 25 ms (from 2.7 s to 2.95 s) long Stored Waveform Inverse Fourier Transform (SWIFT) pulse was applied to one of the end caps of the ion trap to isolate the parent species.\cite{doroshenko1996advanced} Shortly afterward there a re-thermalization period was given for 0.05 s (from 2.95 s. to 3.00 s). It should be noted that the intense electron impact ionization source not only produced parent PAH cations but fragments due to H/\ce{H2} loss as well. For the purposes of this work, the SWIFT pulse was applied to isolate the parent cations as well as the cations corresponding to H/\ce{H2} losses from the parent (i.e., the 247- 252 amu range) and efficiently filter out other spurious signals or contaminations. After the SWIFT pulse was employed, the laser beam shutter was opened, and the ion cloud was irradiated. Typically, 40 pulses and 30 pulses from the dye laser were used for corannulene and sumanene cations, respectively. At the end of each irradiation time with the specified number of pulses, the ions were accelerated out of the ion trap and into the field-free TOF region at the end of which the ions were detected by a multichannel plate (MCP) detector and digitized using a 8-bit GaGe Cobra card. The data acquisition is done using LABVIEW software to obtain and calibrate the mass spectrum. 

For each mass spectrum, an average of 50 scans were used and before each measurement, a normalization spectrum of 15 scans were taken with the laser off condition. These normalization spectra were used to normalize the peak areas of the mass spectra measured with the laser on, which resulted in errors of ~5\%. The number of laser pulses showcased in the experiments are 40 pulses for Corannulene cations and 30 pulses for Sumanene cations, in steps of 2 pulses (e.g.: 0, 2, 4, 6, … 30). These numbers of pulses were chosen such that most of the parent cations would be depleted in the ion trap and converted to fragments. In each case, the total cycle duration was kept constant at 6 s and 7 s for corannulene and sumanene cations respectively to ensure that all datasets are cross comparable and the only changed parameter is pulse energy and/or number of pulses. 

In short, \textit{ion gate opening $\longrightarrow$ trapping ions in the ion trap $\longrightarrow$ mass isolation using SWIFT $\longrightarrow$ photolysis by specified number of laser laser pulses in a squence (0 or 2 or 4...) $\longrightarrow$ extraction of fragment ions $\longrightarrow$ recording the TOF mass spectrum} is a full cycle that repeats for each measurement. The recorded mass spectrum was calibrated from time domain to m/z using a MATLAB code. The same software was also used to deduce the peak are of each cationic peak using a Pearson type IV fit as this gave more reliable result for peak area compared to a gaussian fit. The obtained peak areas were normalized using the normalized scans of mass spectrum as described above, and these normalized values of peak area was used to make the breakdown diagrams presented in this paper. 

\section{Theory}

Density Functional Theory (DFT) was used to explore the potential energy surfaces (PES) of corannulene and sumanene cations. The Minnesota functional M06-2x which includes dispersion in combination with a triple-zeta quality basis set - 6-311++G(3df,2pd) - was chosen, including polarisation and diffuse functions which are needed to describe accurately reaction energies and 
barrier heights with density functional theory \cite{DiffuseFunction}. For example, the energy difference between vinylidene (\ce{CCH2}) and acetylene (\ce{C2H2}) is predicted to be 1.85 eV with this method while is calculated to be 1.86 eV at CCSD(T)/cc-pVTZ level.\cite{chang1997extended}
All the calculations were performed with Gaussian C.02\cite{frisch2016nakatsuji} and the results visualized with the software Molden.\cite{schaftenaar2000molden} Intrinsic Reaction Coordinate (IRC) calculations were performed to check that the transition states connected the intended reactants and products. The expectation value of the total spin of the system \textlangle$S^2$\textrangle was used as proxy of the presence of higher spin states. In the large majority of the cases the structure with the lowest spin multiplicity (e.g. doublet for the isomers of \ce{C20H10+}) is also the more stable structure and that is the one we included in the PES. Special cases are indicated in the discussion. Throughout the paper, energies are zero-point corrected and expressed in eV. Due to the large number of possible photodissociation channels, the exploration of the PES with static DFT calculations is not exhaustive, yet it can shed light on some of the fragmentation behaviours seen in the experiments.

\section{Results and Discussion}
\subsection{Corannulene cation (\ce{C20H10+})}

\begin{figure}[b]
\centering
  \includegraphics[height=5cm]{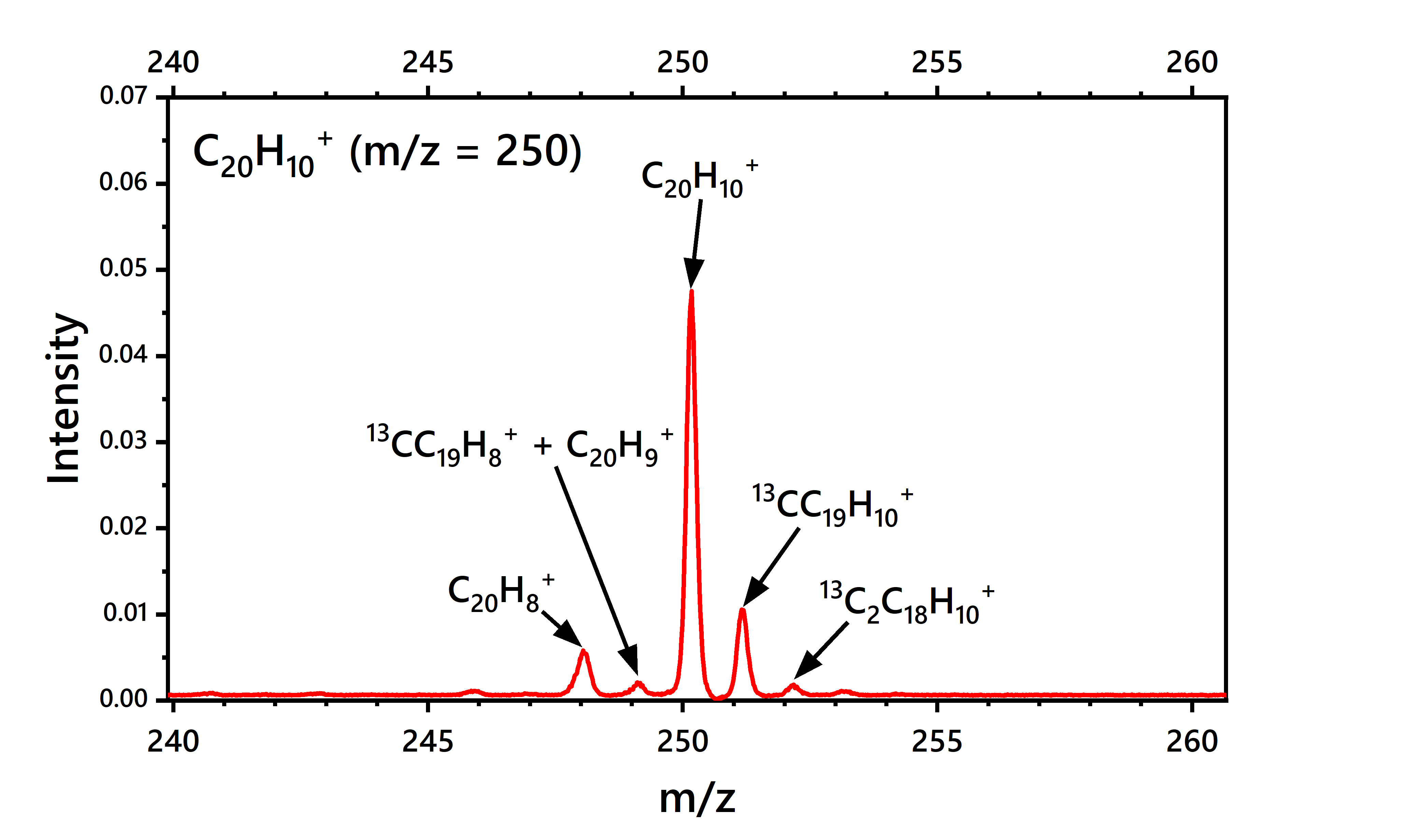}
  \caption{Mass spectrum of \ce{C20H10+} obtained upon electron impact using the Re-TOF instrument integrated in the i-PoP system.}
  \label{Cor_MS}
\end{figure}

\begin{figure*}[h]
    \centering
    \begin{subfigure}{10cm}
        \centering
        \includegraphics[width=\linewidth]{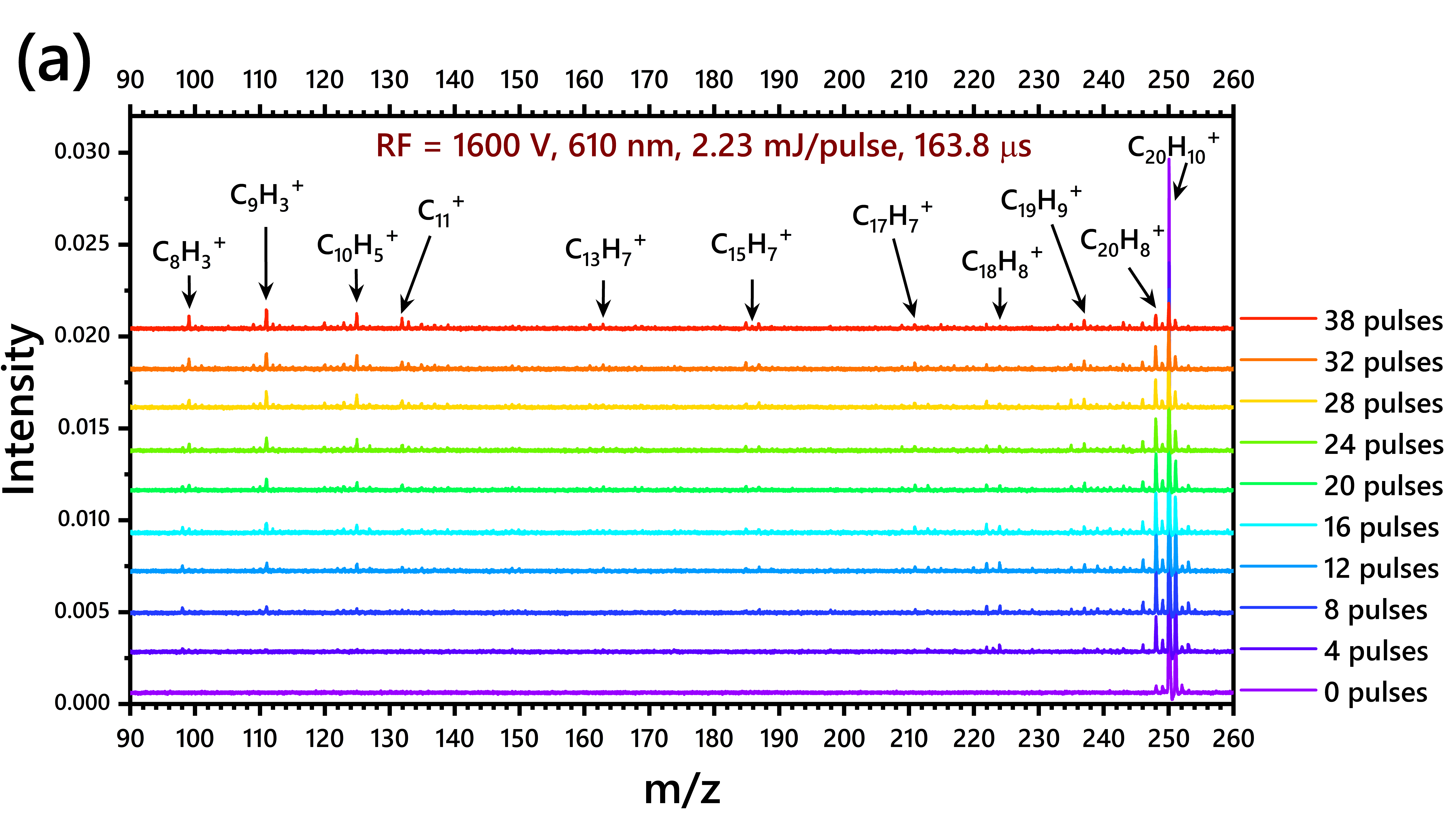}
        \phantomsubcaption
        \label{subfig:1600_cor_MS}
    \end{subfigure}
   \hspace{\columnsep} 
    \begin{subfigure}{6.5cm}
        \centering
        \includegraphics[width=\linewidth]{figures/Figure3b_PCCP.png}
        \phantomsubcaption
        \label{subfig:1600_cor_BD}
    \end{subfigure}
   
    \caption{(a) TOF mass spectra for corannulene radical cations irradiated with 0, 2, 4, … and 40 laser pulses (mass spectra is presented in steps of 4 pulses for clarity) with a pulse energy of 2.23 mJ/pulse with RF = 1600 V i.e. spectra displaying the high-mass cations with prominent features at m/z = 260-140 range; (b) Breakdown diagrams for the fragmentation of \ce{C20H10+}. The normalized mass peak areas as function of the number of pulses are plotted, corresponding to the most prevalent hydrocarbon compounds are formed in with the irradiation with a 610nm laser, with 2.23 mJ/pulse and RF of 1600 V.}
    \label{fig:1600_cor}
\end{figure*}

Figure \ref{Cor_MS} shows the mass spectrum of \ce{C20H10+} obtained with the i-PoP system after applying the SWIFT pulse. The m/z = 250 corresponds to the parent cation \ce{C20H10+}. The less intense peaks at m/z = 251 and 252 are due to the 13C components present in the sample. It is also observed that the electron impact on gaseous \ce{C20H10+} to ionize the neutral parent species also induces H abstraction to produce \ce{C20H9+} and \ce{C20H8+}.
\\
An electron energy of 70 eV and emission of 0.6 mA was used to produce a suitable amount of corannulene cations, \ce{C20H10+}. Figure \ref{subfig:1600_cor_MS} presents the TOF-MS of \ce{C20H10+} for the pulse energies of 2.23 mJ/pulse. The usage of red light with a pulse energy of 2.23 mJ/pulse allows us to decode a number of interesting fragmentation pathways. The fragments produced upon laser irradiation shown in these figures are cationic i.e. neutral fragments, would not be detectable through ion trap TOF mass spectrometry. In these experiments, 50 scans were taken for each mass spectrum. Such mass spectrum were obtained after 2 pulses, 4 pulses, etc. By accumulation of the fragments upon sequential laser induced fragmentation, the mass spectra are stacked to visualize the increase in fragmentation yield, as presented in figure \ref{subfig:1600_cor_MS}. The RF value used for the experiment in Figure \ref{subfig:1600_cor_MS} is 1600 V which is efficient to detect mass peaks from 98 – 260 m/z, which can display the CH/\ce{C2H2} loss channels in good detail.


An electron energy of 70 eV and emission of 0.6 mA was used to produce a suitable amount of corannulene cations, \ce{C20H10+}. Figure \ref{fig:1600_cor} presents the TOF-MS of \ce{C20H10+} for the pulse energies of 2.23 mJ/pulse. The usage of red light with a pulse energy of 2.23 mJ/pulse allows us to decode a number of interesting fragmentation pathways. The fragments produced upon laser irradiation shown in these figures are cationic i.e. neutral fragments, would not be detectable through ion trap TOF mass spectrometry. In these experiments, 50 scans were taken for each mass spectrum. Such mass spectrum were obtained after 2 pulses, 4 pulses, etc. By accumulation of the fragments upon sequential laser induced fragmentation, the mass spectra are stacked to visualize the increase in fragmentation yield, as presented in figure \ref{subfig:1600_cor_MS}. The RF value used for the experiment in Figure \ref{subfig:1600_cor_MS} is 1600 V which is efficient to detect mass peaks from 98 – 260 m/z, which can display the CH/\ce{C2H2} loss channels in good detail. 

\textit{Photodissociation of \ce{C20H10+} - the high mass range:} As known for any PAH, dehydrogenation is the first step in the photo-fragmentation process. \ce{C20H10+} first loses H and \ce{H2}/2H, and it can be observed that the peak corresponding to \ce{C20H8+} is more intense than \ce{C20H9+}.  This means that the second H atom is easily abstracted compared to the first H atom. Likewise, the fourth H atom loss is also observed more prominently than the third H atom loss. The even-numbered favouring for H loss channels were also notice in other PAH cations like coronene (\ce{C24H12+}), Hexabenzocoronene (\ce{C42H18+}) and Dibenzopyrene (\ce{C24H14+})\cite{hrodmarsson2022similarities}  and seems to be independent of the molecular structure. 

However, unlike other medium-sized PAHs, \ce{C20H10+} loses only four H atoms before the carbon skeleton starts opening up through CH loss. In fact, the CH loss channel is one of the most dominant channels for the fragmentation of \ce{C20H10+}. The CH loss fragments \ce{C19H9+}, \ce{C18H8+} and \ce{C17H7+} are clearly observed in the mass spectrum: 
\begin{align*}
\ce{C20H10+} (-\ce{CH}) &\longrightarrow \ce{C19H9+} (-\ce{CH}) \\ \longrightarrow \ce{C18H8+} (-\ce{CH}) &\longrightarrow \ce{C17H7+}. 
\end{align*}

We note that there could be a competition between the CH loss and \ce{C2H2} loss channel as the first \ce{C2H2} loss channel (220 m/z – 228 m/z) is also getting more intense at pulse 12 and then decreases. This means that \ce{C2H2} loss channel is possible from \ce{C20H10+} with either single or a two-step fragmentation:
\begin{align*}
\ce{C20H10+} (-\ce{CH}) \longrightarrow \ce{C19H9+} (-\ce{CH}) \longrightarrow \ce{C18H8+}\\
\ce{C20H10+} (-\ce{C2H2}) \longrightarrow \ce{C18H8+} 
\end{align*}

After three sequential CH loses, i.e. after the parent , C20H10+, becomes \ce{C17H7+}, it is not possible to lose anymore CH. Figure \ref{subfig:1600_cor_BD} demonstrates that after this point the C2H2 loss channel becomes dominant. The consecutive \ce{C2H2} loss can be described as:
\begin{align*}
\ce{C17H7+} (-\ce{C2H2}) \longrightarrow \ce{C15H5+} (-\ce{C2H2}) \longrightarrow \ce{C13H3+}.
\end{align*}

Alternately, it is also possible for \ce{C17H7+} to undergo subsequent loss of two \ce{C2} from the carbon skeleton:
\begin{align*}
\ce{C17H7+} (-\ce{C2}) \longrightarrow \ce{C15H7+} (-\ce{C2}) \longrightarrow \ce{C13H7+}.
\end{align*}

It is noticeable from the mass spectrum (Figure \ref{subfig:1600_cor_MS}) that the cations \ce{C17H7+}, \ce{C15H7+} and \ce{C13H7+} are sequentially produced after the pulses 4, 8 and 12 respectively, which demonstrates that the \ce{C2} loss in probably sequential.

\begin{figure*}[h]
    \centering
    \begin{subfigure}{10cm}
        \centering
        \includegraphics[width=\linewidth]{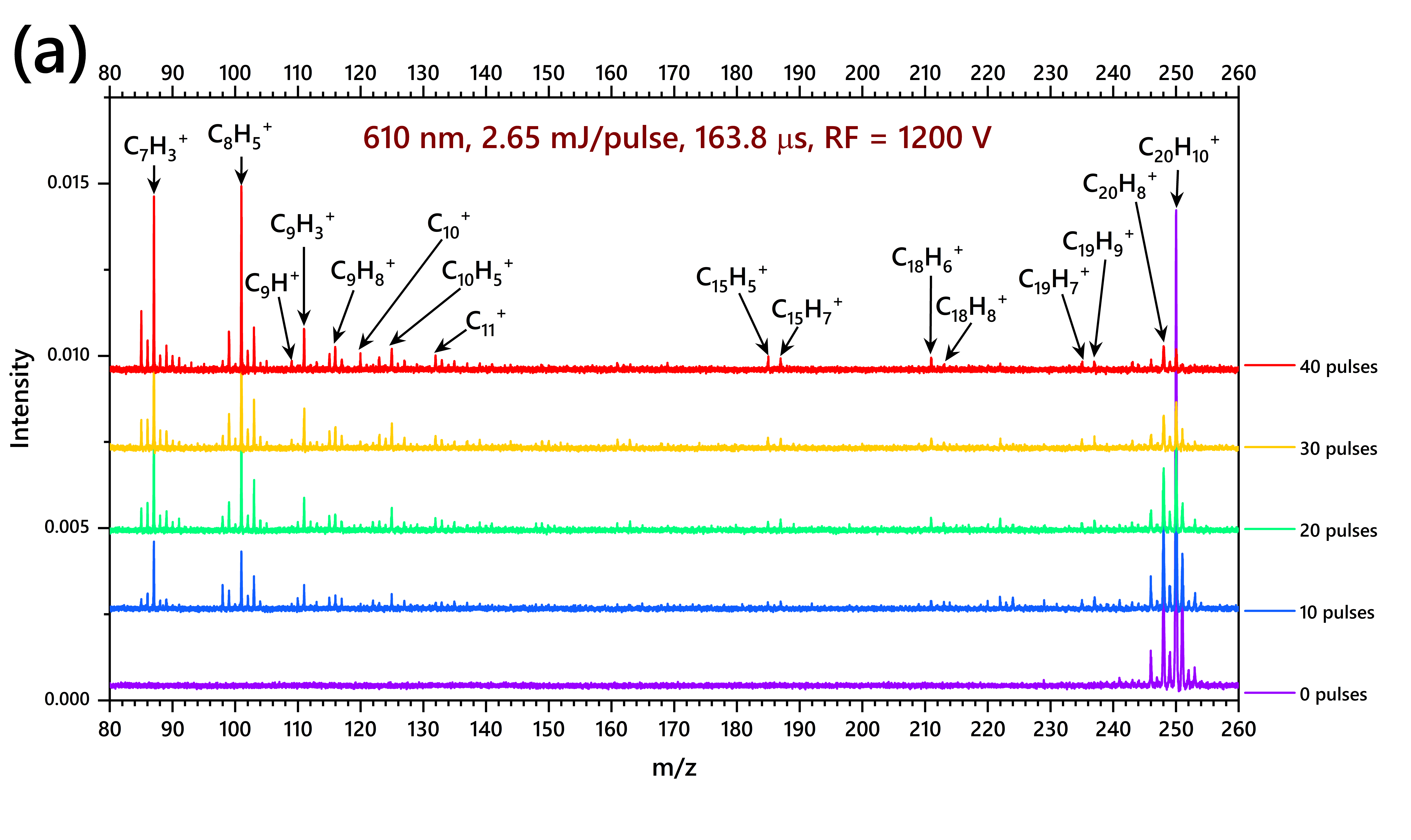}
        \phantomsubcaption
        \label{subfig:1200_cor_MS}
    \end{subfigure}
   \hspace{\columnsep} 
    \begin{subfigure}{6.7cm}
        \centering
        \includegraphics[width=\linewidth]{figures/Figure4b_PCCP.png}
        \phantomsubcaption
        \label{subfig:1200_cor_BD}
    \end{subfigure}
   
    \caption{(a) TOF mass spectra for corannulene radical cations irradiated with 0, 2, 4,… and 40 laser pulses (mass spectra is presented in steps of 10 pulses for clarity) with a pulse energy of 2.65 mJ/pulse with RF = 1200 V i.e. spectra displaying the low-mass cations with prominent features at m/z = 140-80 range; (b) Breakdown diagrams for the fragmentation of \ce{C20H10+}. The normalized mass peak areas as function of the number of pulses are plotted, corresponding to the most prevalent hydrocarbon compounds s formed in with the irradiation with a 610nm laser, with 2.65 mJ/pulse and RF of 1200 V.}
    \label{fig:1200_cor}
\end{figure*}

Figure \ref{subfig:1600_cor_BD} depicts the breakdown diagrams of the parent and product cations: the top panel shows the increase of \ce{C20H8+} upon laser irradiation up to pulse 10 and then a decrease in peak area, which could mean that this species fragments further into smaller secondary cations. Similarly in the middle panel \ce{C18H8+} and \ce{C18H6+} increases up to pulse 10 and then decreases. At this juncture (from pulse 10), the smaller cations \ce{C8H5+}, \ce{C9H5+} and \ce{C11+} start to form and increases steadily. The breakdown equations predicted in this section for the origin of low mass cetions (below 150 m/z) are purely based on the data obtained in the mass spectrum and the breakdown diagram. The breakdown of this corannulene cation is expected to involve several intermediate steps including H roaming, \chemfig{C-C} bond cleavage followed by opening of the C skeleton, etc. before the formation of the observed low mass cations. The breakdown channels are the predicted based on the end products and do not include any intermediate or transition states. So caution must be taken in perceiving the breakdown predictions presented in this paper. 

Possible breakdown mechanisms for \ce{C18H8+} could be:
\begin{align*}
\ce{C20H10+} (-\ce{C2H2}) &\longrightarrow \textbf{\ce{C18H8+}} \longrightarrow \ce{C8H5+} + \ce{C10H3}, \\
\ce{C20H10+} (-\ce{C2H2}) &\longrightarrow \textbf{\ce{C18H8+}} \longrightarrow \ce{C9H5+} + \ce{C9H3}, \\
\ce{C20H10+} (-\ce{C2H2}) &\longrightarrow \textbf{\ce{C18H8+}} \longrightarrow \ce{C11+} + \ce{C7H8}. 
\end{align*}

The \ce{C18H6+} product can be produced after two hydrogen loss followed by two consecutive CH losses, e.g. 
\begin{align*}
\ce{C20H10+} (-\ce{H2}/2H) &\longrightarrow \ce{C20H8+} (-\ce{CH}) \\ &\longrightarrow \ce{C19H7+} (-\ce{CH}) \longrightarrow \ce{C18H6+}. 
\end{align*}

Or with two hydrogen losses followed by a \ce{C2H2} loss to eventually produce \ce{C8H5+}, \ce{C9H5+} and \ce{C11+} with the mechanism: 
\begin{align*}
\ce{C20H10+} (-\ce{H2}) &\longrightarrow \ce{C20H8+} (-\ce{C2H2}) \longrightarrow  \ce{C18H6+} \\ &\longrightarrow \ce{C8H5+} + \ce{C10H};
 \\(or) &\longrightarrow \ce{C9H5+} + \ce{C9H};
 \\(or) &\longrightarrow \ce{C11+} + \ce{C7H6}. 
\end{align*}

It is evident from figure \ref{subfig:1600_cor_BD} that the parent \ce{C20H10+} and its ${^1}$${^3}$C component starts decreasing immediately after the laser pulses are applied; and there are few fragment cations produces right from the beginning of laser irradiation. Nevertheless, there are several fragment cations that are visible only after 6 to 10 laser pulses. This could either be because the cations visible after 6 laser pulses are initially less abundant, or, they are secondary cations (i.e. they are fragmented from the cations after H/CH/\ce{C2H2} loss of \ce{C20H10}) Hence,  these large cations, after CH/C2H2 losses, further fragment into smaller species upon secondary fragmentation to produce the observed low mass cations (C$\leq$11).

\ce{C11+} could also be produced directly from the parent cation upon two or more dehydrogenation of the parent, followed by ring opening and isomerization: 
\begin{align*}
\ce{C20H10+} \longrightarrow \ce{C20H\textsubscript{10-n}+} \longrightarrow \ce{C11+} + \ce{C9H\textsubscript{10-n}}. 
\end{align*}

This is because the radical cites are necessary for the ring opening and isomerization process to take place. Similar to \ce{C11+}, another carbon cation \ce{C10+} is also observed as a secondary fragment only after pulse 16 (Figure \ref{subfig:1600_cor_MS}): 
\begin{align*}
\ce{C20H10+} (-\ce{H2}/2H) &\longrightarrow \ce{C20H8} (-\ce{C2H2}) \longrightarrow \ce{C18H6+} \\ &\longrightarrow \ce{C10+} + \ce{C8H6} \hspace{0.5cm} (or) \\ 
\ce{C18H8+} &\longrightarrow \ce{C18H\textsubscript{8-n}+} \longrightarrow \ce{C10+} + \ce{C8H\textsubscript{8-n}}. 
\end{align*}

The other intense low mass fragments like \ce{C8H3+}, \ce{C8H5+}, \ce{C9H3+} and \ce{C10H5+} are observed from the beginning of laser irradiation and are observed to be primary cationic fragments from the parent cation immediately after H loss: 
\begin{align*}
\ce{C20H10+} &\longrightarrow \ce{C20H\textsubscript{10-n}+} \longrightarrow \ce{C9H3+} + \ce{C11H\textsubscript{7-n}} \\
\ce{C20H10+} &\longrightarrow \ce{C20H\textsubscript{10-n}+} \longrightarrow \ce{C10H5+} + \ce{C10H\textsubscript{5-n}}
\end{align*}

\textit{Photodissociation of \ce{C20H10+} - the low mass range:} Figure \ref{subfig:1200_cor_MS} shows the mass spectrum upon irradiation with sequential laser pulses using RF = 1200 V, illustrating a striking difference in the nuances of the low mass region (80 – 130 m/z) compared to figure \ref{subfig:1600_cor_MS} (i.e. experiment with RF=1600) because of the reduced RF value to 1200 V in this case. However, the CH and \ce{C2H2} losses that are prominently observed in the experiment with RF = 1600 V, are also observed in figure \ref{subfig:1200_cor_MS}.\cite{march2005quadrupole} In this case, smaller photofragments like \ce{C7H3+}, \ce{C8H5+}, \ce{C9H3+}, \ce{C10+} and \ce{C11+} are observed in higher intensity. We specifically note that in figure \ref{subfig:1600_cor_BD}, \ce{C8H3+} dominates over \ce{C8H5+}, while in figure \ref{subfig:1200_cor_BD}, \ce{C8H5+} is dominant over \ce{C8H3+}. We consider it possible that, besides \ce{C8H5+}, the 101 m/z peak has a contribution by the \ce{C7H+}-\ce{H2O} reaction that gives rise to \ce{C7HO+} (or do we mean \ce{C7H+} + \ce{H2O} product) due to trace contamination in the ion chamber and further experiments are underway at our laboratory to investigate this further. For that reason, we refrain from a further analysis of the low mass range

It is evident from the breakdown diagrams in Figure \ref{subfig:1200_cor_BD} (right panel) \ce{C7H3+} and \ce{C8H5+} are formed from the very beginning of laser irradiation, e.g. 
\begin{align*}
\ce{C20H10+} \longrightarrow \ce{C7H3+} + \ce{C13H7} \\
\ce{C20H10+} \longrightarrow \ce{C8H5+} + \ce{C12H5}
\end{align*}

Again, \ce{C10+} and \ce{C11+} appears to be secondary products arising from pulse 10 and pulse 6 respectively (Figure \ref{subfig:1200_cor_BD}). The other cationic products are in relatively lower abundance and are also produces right after the irradiation with first two laser pulses. This could mean that an energy of ~5.3 mJ is already sufficient to break the carbon skeleton of \ce{C20H10+} into two or more fragments.  

\subsubsection{Theoretical results for the fragmentation of Corannulene cation}

\begin{figure}[h]
\centering
  \includegraphics[height=7cm]{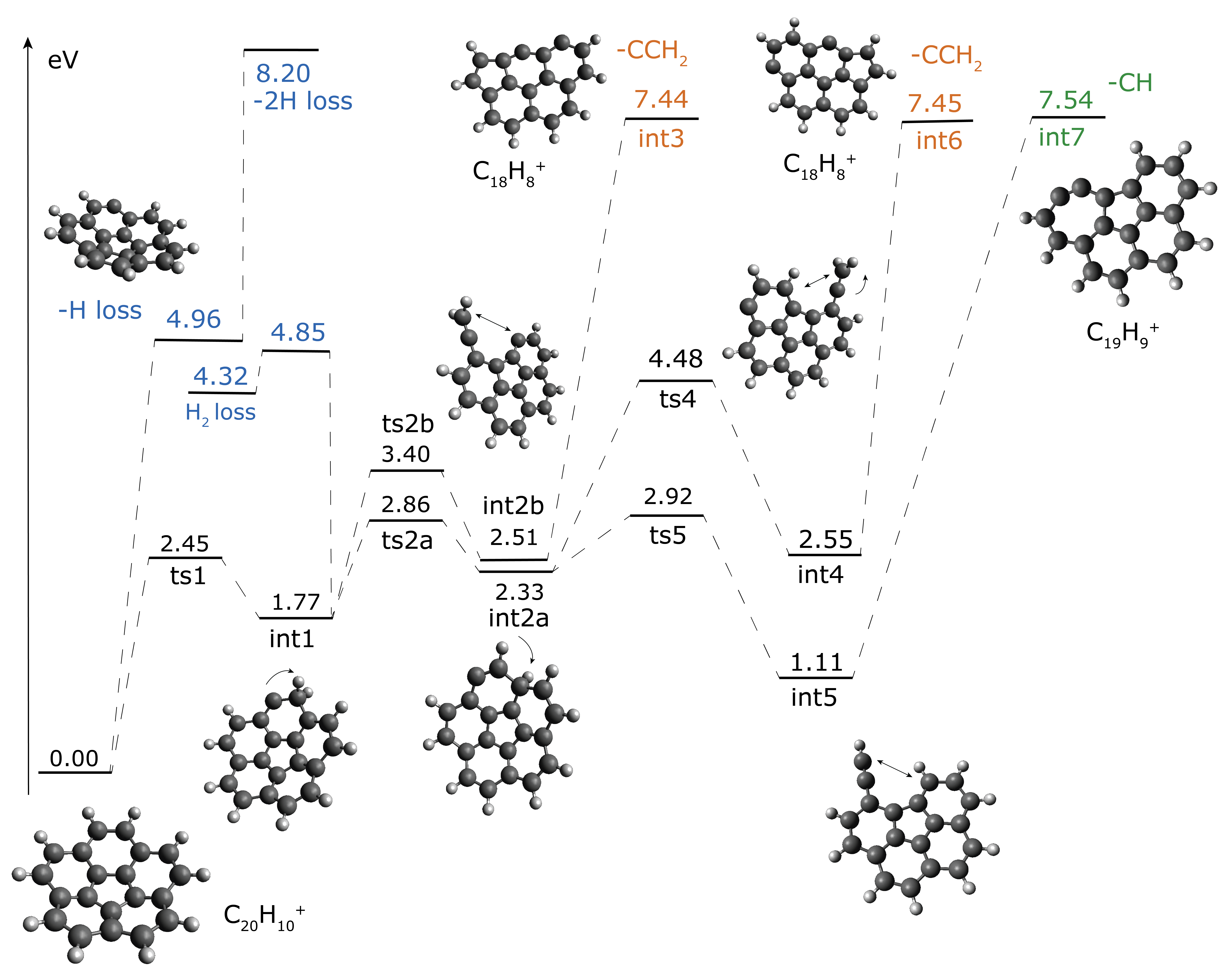}
  \caption{Potential Energy Surface (PES) for the \ce{C20H10+}, obtained at M06-2X/6-311++G(3df,2pd) level. Energies are given relative to the energy of the corannulene cation. Channels are colour-coded to help visualization. All the structures in the PES are at their lowest multiplicity, except for \ce{C20H9+} (triplet), \textit{int3} and \textit{int6} (quadruplet)}
  \label{Cor_PES1}
\end{figure}

In this section the results from the DFT calculations on the corannulene cation and its relevance with the experimental results are discussed The first focus is on the H/2H loss and CH/\ce{C2H2} loss, which will then proceed to proposing some pathways for the further loss of CH and \ce{C2H2}, observed in the experiments. 
\vspace{0.5cm}
\textit{H/2H/\ce{H2}}, CH and \ce{CCH2} loss channels: Figure \ref{Cor_PES1} shows the potential energy surface for the corannulene cation which includes proposed mechanisms leading to loss of H/2H/\ce{H2}, CH and \ce{C2H2}. 

The lowest dissociation product is the dehydrogenated corannulene via H loss, that can happen directly from the corannulene cation with bond dissociation energy (BDE) of 4.96 eV or after isomerization reactions involving H roaming.\footnote{In the rest of the paper we use the term H roaming and H migration interchangeably.} The H roaming reactions require up to 4 eV, and each of the intermediates created (\textit{int1}, \textit{Int2a} and \textit{int2b}, Figure \ref{Cor_PES1}) can lose an H atom, all of which leads to the same final structure, i.e. singly-dehydrogenated curved corannulene (\ce{C20H9+}). The BDEs for the H-shifted intermediate can be derived from the difference in energy between corannulene and singly dehydrogenated corannulene and are: \vspace{0.2cm} \\ 
\vspace{0.2cm}
3.19 eV for \textit{int1} (loss from the aliphatic group), \\
\vspace{0.2cm}
2.63 eV for \textit{int2a} (loss from the tertiary carbon) and \\
\vspace{0.2cm}
2.45 eV for \textit{int2b} (loss from the vinylidene \ce{CCH2} chain).
\vspace{0.3cm}

H loss from \textit{int4} can proceed from the \ce{CCH2} chain at slightly higher BDE (3.28 eV) than for the other H-shifted isomers; this is because, in this case losing one of the aliphatic hydrogens from the chain will create a high-lying isomer of \ce{C20H9+} rather than the more stable singly-dehydrogenated corannulene structure. For \textit{int5}, all the H atoms present in the molecule are aromatic (Figure \ref{Cor_PES1}) and their BDE will be around 5 eV. The loss of a second H atom from the singly-dehydrogenated corannulene cation has a BDE of 3.24 eV. We also investigated the \ce{H2}-loss channel from \textit{int1}. The transition state is at 4.85 eV above corannulene. This behaviour and BDEs presented in this work agrees very well with previous studies on planar PAHs.\cite{D4CP01301H}$^,$\cite{castellanos2018photoinduced,trinquier2017pah,west2014photodissociation}
\\
Isomerization is also the first step toward the C-loss channels, which needs comparable energies of 7.4-7.5 eV (Figure \ref{Cor_PES1}). Once \textit{int1} is created, the bond between two carbon atoms can be broken with a transition state (\textit{ts2b}) of 3.40 eV, leading to the formation of a vinylidene group (\textit{int2b}, 2.51 eV) on an almost flat molecule. From there a \ce{CCH2} unit can be released through a \chemfig{C-C} bond dissociation with 7.44 eV of energy with respect to corannulene (4.93 eV from \textit{int2b}). Interestingly, analysis of the optimized structures along the \chemfig{C-C} dissociation curve shows that the loss of a \ce{CCH2} unit triggers the rearrangement of 2 carbon bonds in the molecule. This leads to a change in the position of the pentagon in the structure, which is now on the outside and thus leads to an almost flat structure for the \ce{C18H8+} isomer. An additional path leading to the release of a \ce{CCH2} unit goes through intermediate \textit{int2a} (2.33 eV) where the H moves to the tertiary carbon atom. The breaking of a \chemfig{C-C} bond and migration of the H atom (\textit{ts4}) leads to the formation of another \ce{C20H10+} isomer with a vinylidene group (\textit{int4})\textcolor{blue}{\footnote{Optimising \textit{int4} with quadruplet multiplicity leads to an alternative structure with a -CH-CH group at 4.24 eV with respect to the corannulene  (1.69 eV higher than int4 in doublet spin state (not in the Figure). From there we found a transition state leading to the loss of \ce{C2H2} with a barrier of 6.36 eV. However the IRC calculation was inconclusive and the products (\ce{C18H8+} + \ce{C2H2}) lies at 7 eV, higher than the transition state, making this path doubtful.}} at 2.55 eV with respect to \ce{C20H10+}.  With additional 4.90 eV, \textit{int4} can lose a \ce{CCH2} unit while undergoing a carbon rearrangement similar to what observed for \ce{CCH2} loss from \textit{int2b}.

The formation of \textit{int2a} can also lead to the creation of an ethynyl group (\textit{int5}) through a lower transition state (\textit{ts5}, 2.92 eV). \textit{Int5} is only 1.11 eV above the corannulene cation. From there, CH can be released from the molecules with 6.44 eV, leading to the formation of an almost flat isomer of \ce{C19H9+} that contains a seven-membered ring. As observed for other intermediates analysed before, the structure rearrangement (the formation of a seven-memberd ring from the insertion of the remaining C atom into the six-membered ring, leading to \textit{int7}) happens toward the last stages of the CH bond cleavage. Calculations show that the structure with a 7-5 membered ring is thermodynamically preferred over a 6-6 structure because the heptagon facilitates the formation of a triple bond — r=1.23 Å — between the two dehydrogenated carbon atoms in the 7-membered ring. 
\\
\\
\begin{figure}[h]
\centering
  \includegraphics[height=5.8cm]{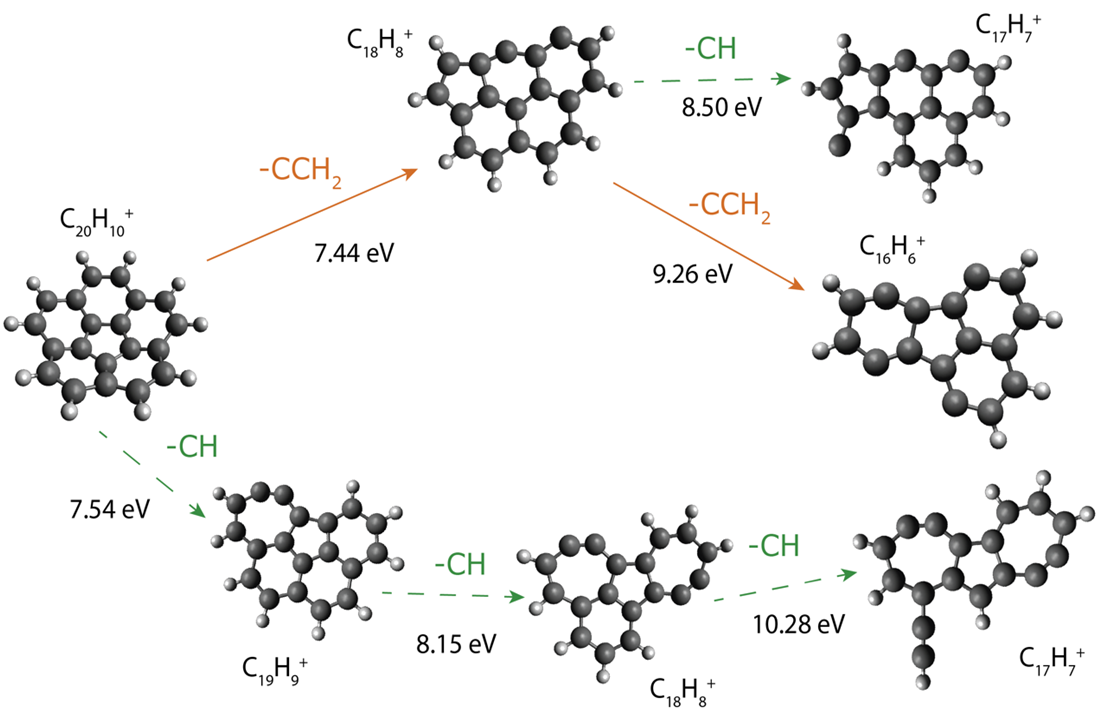}
  \caption{DFT optimized structures and reaction energies involved in the sequential loss of 2 \ce{CCH2} (orange solid lines), \ce{CCH2} and CH, and 3 CH loss (green dashed lines) from the corannulene cation. }
  \label{Cor_PES2}
\end{figure}

\textit{Further CH and \ce{C2H2} losses:} Figure \ref{Cor_PES2} shows two possible pathways toward further \ce{CCH2} and CH loss from the daughter molecules of corannulene cations, namely \ce{C19H9+}, \ce{C18H8+} and \ce{C17H7+}, informed from the results obtained for the fragmentation of parent molecule. Since the limiting step for both H- and C-loss channels in corannulene cation is the direct dissociation of a \chemfig{C-H} or \chemfig{C-C} bond, we investigate here only the final structures and not the H-roaming/\chemfig{C-C} breaking steps leading to fragmentation. Rearrangements of carbon structure, even if likely happening, are not considered here. As explained above, the corannulene cation can lose a \ce{CCH2} group, leading to the formation of a \ce{C18H8+} isomer (orange solid arrow, top part of Figure \ref{Cor_PES2}). From there, H roaming isomerization, similar to what is observed in corannulene, promotes the opening of the bottom left hexagonal ring (see Figure \ref{Cor_PES1}) and can lead to \ce{C17H7+} thanks to a CH loss (with a BDE of 8.50 eV) or to \ce{C16H6+} through a \ce{CCH2} (BDE of 9.26 eV). It is expected that this set of reactions can proceed also from the bottom right ring of \ce{C18H8+} with similar energetics. For \ce{C18H8+} the BDEs for CH and \ce{CCH2} have a larger difference (0.76 eV) than for the corannulene cation (0.10 eV). It is interesting to notice that the loss of CH from \ce{C18H8+} does not promote the rearrangement of the pentagon into a hexagon but requires an additional reaction step. The isomer of \ce{C17H7+} considered here already possess two isolated radical sides and likely the formation of a new bond due to rearrangement of the pentagon-C group into a hexagon does not lower much the energy of the molecule.

The bottom sequence of figure \ref{Cor_PES1} illustrates the structures related to sequential loss of 3 CH groups from \ce{C20H10+} (green dashed arrow). Starting from \ce{C19H9+}, a CH loss from the bottom right hexagon will lead to the formation of a \ce{C18H8+} isomer with two heptagons. This requires 8.15 eV of energy, just 0.7 eV higher that the BDE for the first CH loss. A third CH loss can proceed from the only hexagon left in \ce{C18H8+} with 10.28 eV, leading to the formation of a \ce{C17H7+} isomer (bottom of Figure \ref{Cor_PES2}). The energy needed for this last CH loss is 2 eV higher than for previous steps: this is likely because the ethynyl chain on \ce{C17H7+} cannot be inserted into the heptagon to create a 9-membered ring that stabilizes the structure.

\subsubsection{Discussion - fragmentation of Corannulene cation}

The experiments on corannulene fragmentations show H and 2H/\ce{H2} loss as the lowest dissociation channel (Figure \ref{Cor_MS} and Figure \ref{subfig:1600_cor_MS}). Looking at the PES, the H loss channel very likely happens from H-shifted isomers of \ce{C20H10+} such as \textit{int1}, \textit{int2a} and \textit{int2b} rather than directly from the corannulene structure. The \ce{H2} loss can also happen from isomer \textit{int1} and the height of the barrier is comparable to that of H-loss. It is not possible, based on energetics only, to constrain if the peak at \textit{m/z=128} is due to sequential hydrogen loss or \ce{H2} loss. Monte Carlo simulation\cite{castellanos2018photoinduced} based on DFT-derived RRKM rates can be used to disentangle between the two contributions but is out of the scope of this paper.

Regarding the C-loss channels, the DFT calculations show that the \ce{C18H8+} peak results from \ce{CCH2} rather than acetylene loss. The investigation of the PES did not lead to any viable routes toward direct or indirect acetylene loss, even considering higher multiplicities See Supplementary information (figure S1) and footnote in the pregiosu section. The competition between the CH and \ce{CCH2} channels deduced from the experiments can be explained looking at the Figure \ref{Cor_PES1}, where the two channels go through \textit{int1} and have similar reaction energies. At lower internal energy the \ce{CCH2} loss happens likely from \textit{int2b}, while the CH loss starts from \textit{int5}. The relative importance of the two channels at different internal energies will depends on the number of possible channels (including back-reactions and further isomerization) and also on the enthalpy of formation $\Delta$S$^\dag$$_{1000}$ of the two \chemfig{C-H} cleavage reactions, that influences the shape of the reaction rate \cite{Baer1997}. Moreover, The irradiation of corannulene cations with the dye laser produces nanosecond pulses leading to a multi-photon excitation process that eventually leads to fragmentation of the molecule involving virtual energy state(s). As the laser pulses are applied the internal energy of the molecular cation increases which leads to fragmentation following the potential energy surface. This being a non-linear photochemical process, both CH and \ce{CCH2} loss channels are produced in competition in spite of the minor energy difference of 1.5 eV (\textit{int 3} and \textit{int 7} compared to \textit{int 2a} and \textit{int 5} respectively) after the H roaming and opening of the C skeleton.

Considering further fragmentation, Figure \ref{Cor_PES2} shows that the sequential loss of two CH units or that of one \ce{CCH2} from \ce{C20H10+} leads to a population of different \ce{C18H8+} isomers. From there only CH loss is observed, leading to \ce{C17H7+}, while there is no trace of \ce{C16H6+} peaks. This is different from what is seen on corannulene cation, where the 2 channels are in competition. Calculations of the reaction energies for CH and \ce{CCH2} loss (Figure \ref{Cor_PES2} top sequence), shows that CH-loss of \ce{C18H8+} requires indeed lower energy. The presence of partly dehydrogenated sites in different rings affect the energy of this reaction, because it prevents rearrangement of the structure leading to isomers with lower energy. As mentioned above, there might be a mixture of \ce{C18H8+} isomers and they will likely not follow the same photodissociation pathway. Also, we cannot exclude that additional isomerization (e.g. H-migration) might take place on some or all of these structures. The presence of two groups of \ce{C18H8+} isomers, behaving differently, could be the reason why the experiments shows that this peak increases upon increasing the laser pulses and starts decreasing on further irradiation (Figure \ref{subfig:1600_cor_MS}). The isomer of \ce{C18H8+}, formed by two consecutive CH losses (as shown in the bottom part of figure \ref{Cor_PES2}) is expected to undergo further fragmentation to form \ce{C11+}. This behaviour could be inferred from figure \ref{subfig:1600_cor_BD} and figure \ref{subfig:1200_cor_BD}, by the fact that the intensity of \ce{C18H8+} trace does not completely drop to zero. 

Finally, the mass spectra obtained with RF value of 1200 V (e.g. Figure \ref{subfig:1200_cor_MS}) show the presence of strong peaks corresponding to \ce{C7H3+} and \ce{C8H5+}. These peaks appear very early in the fragmentation process, and we have postulated they might come directly from the parent molecule, likely after isomerization reaction. Indeed, it has been proposed that carbon skeleton rearrangement (e.g. formation of 7-membered rings) in PAHs can lead to dissociation to long hydrocarbon chains or carbon rings.\textcolor{blue}{\cite{D4CP01301H}} Looking at Figure \ref{Cor_PES2}, we note that the \ce{C17H7+} isomer at the bottom of the figure, and its parent molecule \ce{C18H8+}, have a 7-membered ring with 3 H atoms. Cleavage of 2 carbon bonds could release of a \ce{C7H3} fragment from the molecules, e.g. (without considering the charge):
\begin{align*}
\ce{C18H8+} \longrightarrow \ce{C7H3+} + \ce{C11H5} \\
\ce{C17H7+} \longrightarrow \ce{C7H3+} + \ce{C10H4}
\end{align*}

In this type of experiment, it is expected that the charge of the parent molecule would remain with the largest fragment, yet a strong peak is observed for \ce{C7H3+} but not for \ce{C10H4+} or \ce{C11H5+}. A possible explanation is that the large fragments mentioned above quickly breaks done in smaller neutral hydrocarbons in the release of \ce{C7H3+}. Molecular dynamics simulations based on DFTB on PAHs have shown these multiple hydrocarbon fragmentation channels if enough energy is available.\cite{simon2017dissociation} It could also be the case that the ionization potential of \ce{C7H3} is smaller than the remaining larger fragment. A previous study on both the cation and radical neutral version of \ce{C7H3} shows that, among the possible isomers, a three-membered carbon ring fused with the linear \ce{C4H} chain is the local minimum of the PES  and its ionization energy is only 1.6eV.

A similar reasoning can be applied to the other abundant hydrocarbon, \ce{C8H5+} that can also be released from \ce{C18H8+} (Figure \ref{Cor_PES2}, bottom) after consecutive cleavage of 3 \chemfig{C-C} bonds, leaving behind a bycyclic structure made of one heptagon and one hexagon. Unfortunately, there are no studies on the properties of \ce{C8H5+} and a detailed investigation of its PES\cite{chakraborty2014spectroscopic} is beyond the scope of this paper.

\subsection{Sumanene cation (\ce{C21H12+})}

\begin{figure}[h]
\centering
  \includegraphics[height=5cm]{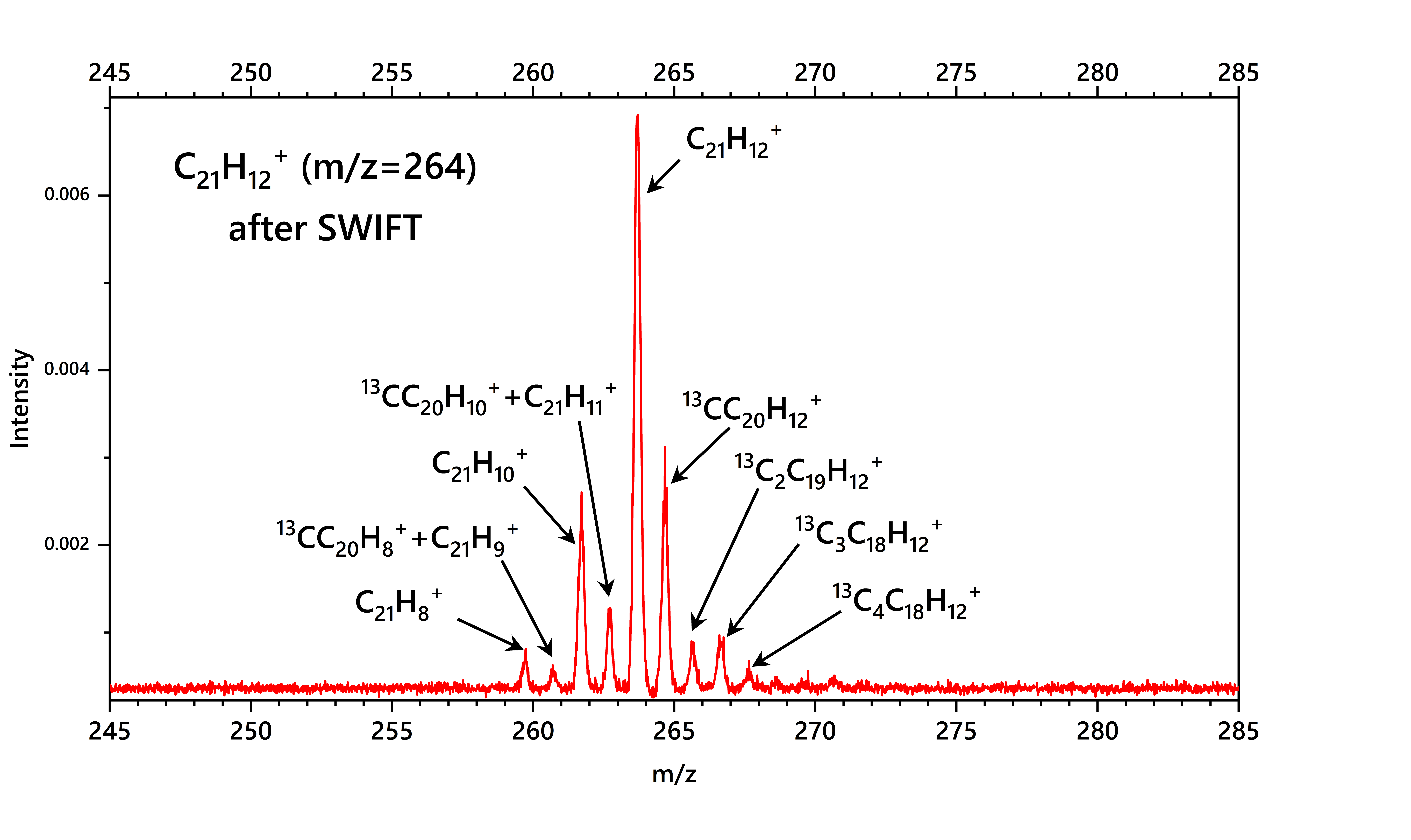}
  \caption{Mass spectrum of \ce{C21H12+} obtained upon electron impact using the Re-TOF instrument integrated in the i-PoP system.}
  \label{Sum_MS}
\end{figure}

An electron energy of 70 eV and emission of 0.6 mA was used to produce good amount of sumanene cations, \ce{C21H12+}. Figure \ref{Sum_MS} shows the mass spectrum of \ce{C21H12+} obtained with the i-PoP system. The m/z = 264 corresponds to the parent cation \ce{C21H12+}. The moderately intense peaks at m/z = 265, 266, 267 and 268 are due to the $^{13}$C components present in the sample. The $^{13}$C isotopic abundance in the sample was estimated to be ~20 \% by estimating their integrated band area with respect to the parent cation. It is to be noted that the natural abundance of $^{13}$\ce{C20H12} is slightly higher than that of other PAHs like corannulene or coronene. The reason behind this is still unknown as sumanene of one of the least studied PAHs till date. However, the abundance of the $^{13}$C components were reduced further ~15 – 17 \% in the photolysis experiments presented in this work by applying a narrow bandwidth SWIFT frequency. But caution must be taken in understanding the fragment cations after laser irradiation as the $^{13}$C isotopes could also contribute to it. It is also observed that the electron impact on gaseous \ce{C21H12+} to ionize the neutral parent species also induces H abstraction to produce \ce{C21H11+}, \ce{C21H10+}, \ce{C21H9+} and \ce{C21H8+}. The mass 263 and 261 could also arise due to the dehydrogenation of the $^{13}$C components producing $^{13}$C\ce{C20H10+} and $^{13}$C\ce{C20H8+} respectively. It is also observed that the double dehydrogenation of the sumanene cation is more favourable by comparing the peak intensity of \ce{C21H11+} and \ce{C21H10+}.

\begin{figure*}[h]
    \centering
    \begin{subfigure}{10.5cm}
        \centering
        \includegraphics[width=\linewidth]{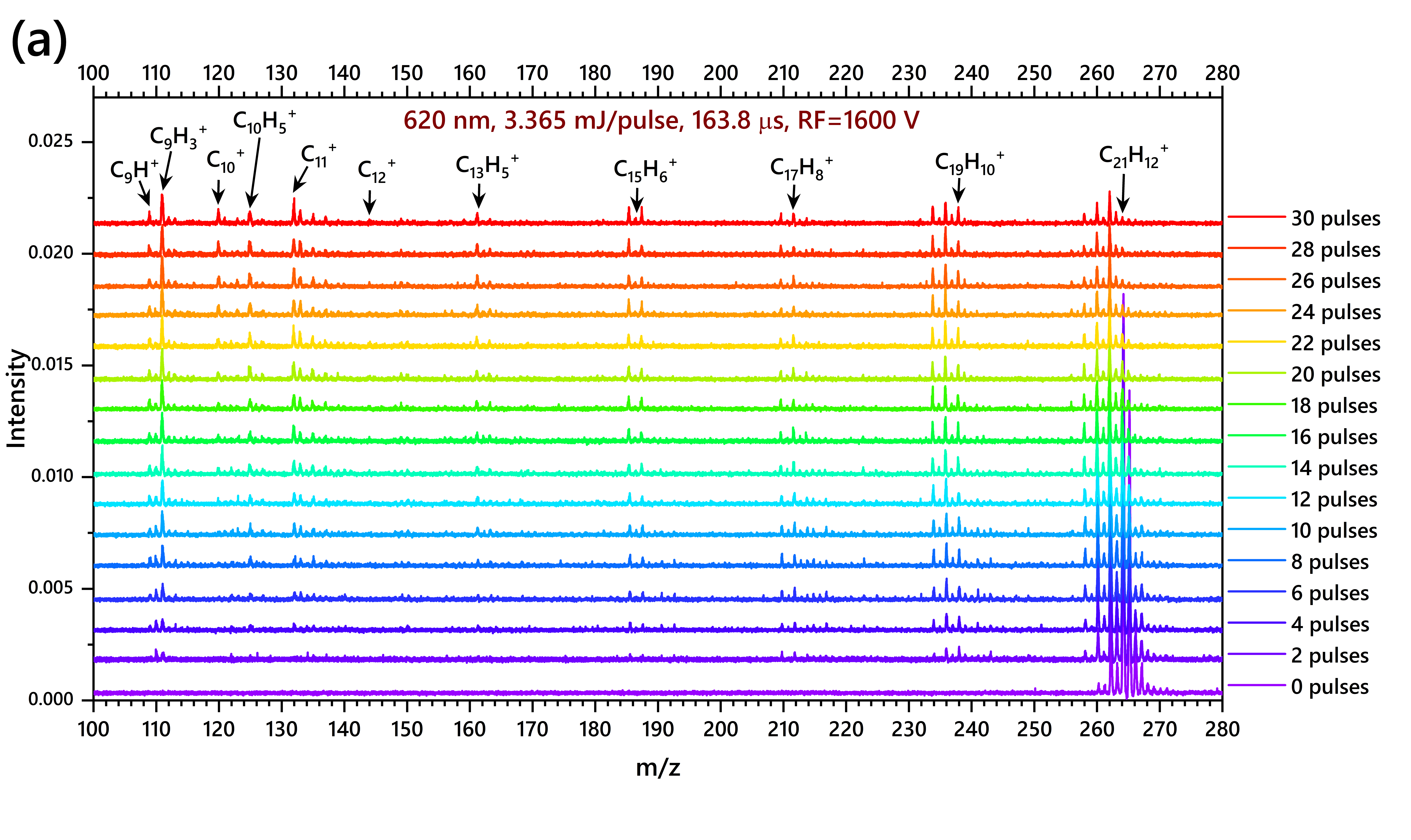}
        \phantomsubcaption
        \label{subfig:1600_sum_MS}
    \end{subfigure}
   \hspace{\columnsep} 
    \begin{subfigure}{6.7cm}
        \centering
        \includegraphics[width=\linewidth]{figures/Figure8b_PCCP.png}
        \phantomsubcaption
        \label{subfig:1600_sum_BD}
    \end{subfigure}
   
    \caption{(a) TOF mass spectra for sumanene radical cations irradiated with 0, 2, 4,… and 30 laser pulses with a pulse energy of 3.365 mJ/pulse with RF = 1600 V i.e. spectra displaying the high-mass cations with prominent features at m/z = 280-150 range; (b) Normalized mass peak areas corresponding to the most prevalent hydrocarbon clusters formed in the fragmentation of the sumanene with RF = 1600 V.}
    \label{fig:1600_sum}
\end{figure*}
\vspace{0.5cm}

\textit{Photodissociation of \ce{C21H12+} - the high mass range:} Figure \ref{subfig:1600_sum_MS} presents the TOF-MS of \ce{C21H12+} for the pulse energies of 3.365 mJ/pulse at 620 nm. The mass spectrum recorded at 0 pulse shows that -H and -2H loss peaks of the parent cation of sumanene is already present upon electron impact. There are also peaks of -3H and -4H present in trace amounts. But as the laser irradiation continues, these H loss peaks increase till 4 pulses and then start to decrease for form secondary cationic fragments as shown in the Figure \ref{subfig:1600_sum_MS} for the case of \ce{C21H10+} and \ce{C21H9+}. In total, six consecutive H loss channels are observed for the parent cation. Sumanene does not completely dehydrogenate unlike Hexabenzocoronene (\ce{C42H18+})\cite{zhen2014laboratory,zhen2014quadrupole} or dibenzopyrene (\ce{C24H14+})\cite{hrodmarsson2022similarities} that consecutively lost all H atoms.

In these experiments, 50 scans were taken for each mass spectrum. Such mass spectrum were obtained after 2 pulses, 4 pulses, etc. By accumulation of the fragments upon sequential laser induced fragmentation of sumanene cation, the mass spectra are stacked to visualize the increase in fragmentation yield, as presented in figure \ref{subfig:1600_sum_MS}. The RF value used in Figure \ref{subfig:1600_sum_MS} is 1600 V which is efficient to detect mass peaks from 98 – 260 m/z. This can display the CH/\ce{C2H2} loss channels in better detail than the low mass cations. It is obvious from the mass spectrum that the \ce{C2H2} is the most dominant channel in \ce{C21H12+} fragmentation. 
\begin{align*}
\ce{C21H12+} (-\ce{C2H2}) &\longrightarrow \ce{C19H10+} (-\ce{C2H2}) \longrightarrow \ce{C17H8+} \\ 
\ce{C17H10+} (-\ce{H}) \hspace{0.2cm}\&\hspace{0.2cm} (-\ce{C2H2}) &\longrightarrow \ce{C15H7+} (-\ce{C2H2}) \longrightarrow \ce{C13H5+} 
\end{align*}

It is to be noted that the first two \ce{C2H2} loses have even number of H atoms and the next two \ce{C2H2} loses have odd number of H atoms. This could be explained by the isomerization of \ce{C21H12+} after two consecutive \ce{C2H2} lose channels, as isomerization becomes the key process in the formation of such rich carbon species especially the low mass cations with m/z < 140. A carbon loss channel could also be expected here, i.e. 
\begin{align*}
\ce{C17H8+} (-\ce{C2}) &\longrightarrow \ce{C15H8+} \hspace{0.4cm} and \\
\ce{C15H5+} (-\ce{C2}) &\longrightarrow \ce{C13H5+}. 
\end{align*}
The \ce{C17H8+} that formed can either undergo further \ce{C2H2} loss and/or a \ce{H2} loss as shown below: 
\begin{align*}
\ce{C17H8+} (-\ce{C2H2}) \longrightarrow \ce{C15H6+} (-\ce{C2H2}) \longrightarrow \ce{C13H4+} \\ (and/or) \hspace{0.5cm}
\ce{C17H8+} (-\ce{H2}) \longrightarrow \ce{C17H6+}
\end{align*}

The \ce{C19H8+} cation which is observed in the mass spectrum can also be formed from the parent cation by two hydrogen losses followed by a \ce{C2H2} loss: 
\begin{align*}
\ce{C21H12+} (-\ce{H2}/\ce{2H}) \longrightarrow \ce{C21H10+} (-\ce{C2H2}) \longrightarrow \ce{C19H8+}. 
\end{align*}
The \ce{C19H8+} is further found to fragment in two possible ways: 
\begin{align*}
\ce{C19H8+} (-\ce{C2H2}) &\longrightarrow \ce{C17H6+} \hspace{0.5cm}(and/or) \\ \ce{C19H8+} (-\ce{H2}) &\longrightarrow \ce{C19H6+}.
\end{align*}

\begin{figure*}[h]
    \centering
    \begin{subfigure}{11cm}
        \centering
        \includegraphics[width=\linewidth]{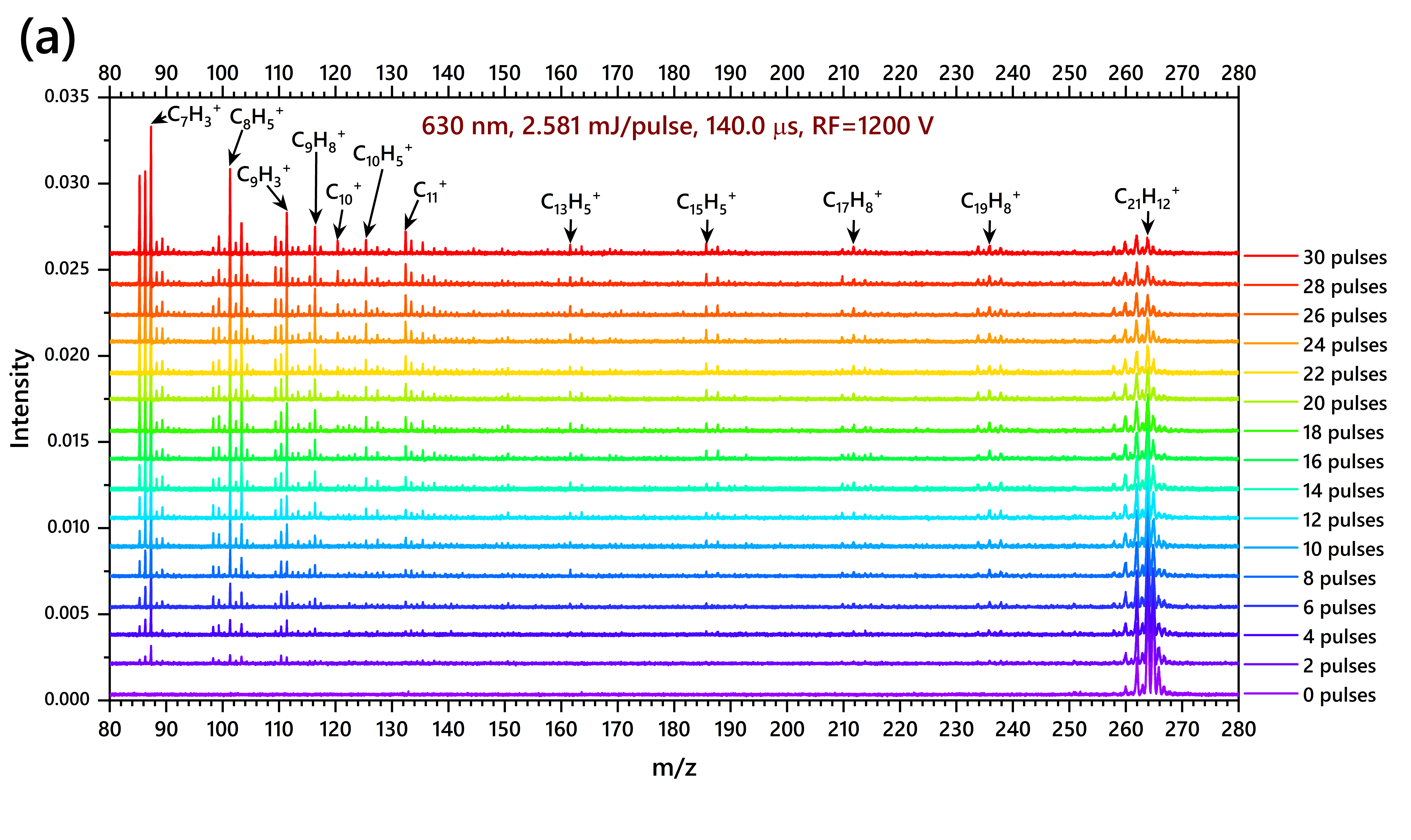}
        \phantomsubcaption
        \label{subfig:1200_sum_MS}
    \end{subfigure}
   \hspace{\columnsep} 
    \begin{subfigure}{6.7cm}
        \centering
        \includegraphics[width=\linewidth]{figures/Figure9b_PCCP.png}
        \phantomsubcaption
        \label{subfig:1200_sum_BD}
    \end{subfigure}
   
    \caption{(a) TOF mass spectra for sumanene radical cations irradiated with 0, 2, 4,… and 30 laser pulses with a pulse energy of 2.581 mJ/pulse with RF = 1200 V i.e. spectra displaying the low-mass cations with prominent features at m/z = 80-150 range; (b) Normalized mass peak areas corresponding to the most prevalent hydrocarbon clusters formed in the fragmentation of the sumanene with RF = 1200 V.}
    \label{fig:1200_sum}
\end{figure*}

\textit{Photodissociation of \ce{C21H12+} - the low mass range:} In the low mass region \ce{C9H3+} is the most abundant product and happens to be one of the primary products of fragmentation, which is also noticeable from the breakdown diagram in Figure \ref{subfig:1600_sum_BD}.

\ce{C10H5+} and \ce{C11+} are the second most abundant products. To get better nuances in this region experiments were performed with lower RF value of 1200 V, as shown in Figure \ref{subfig:1200_sum_MS}, where the masses below 140 m/z are intense. Here \ce{C7H3+}, \ce{C8H5+} and \ce{C9H3+} are the most abundant products of fragmentation of \ce{C21H12+}, which is followed by \ce{C9H8+}, \ce{C10H5+} and \ce{C11+}. It is evident from Figure \ref{subfig:1200_sum_BD} that these low mass cations are formed from the very beginning of laser irradiation. However, the breakdown equations predicted in this section for the origin of low mass cations (below 150 m/z) are only based on the data obtained in the mass spectrum and the breakdown diagram just like the case of corannulene. The intermediate ions and transition states before the formation of the observed low mass cations are not showcased in the pathways given below. 
\begin{align*}
\ce{C21H12+} &\longrightarrow \ce{C21H\textsubscript{12-n}+} \longrightarrow \ce{C7H3+} + \ce{C14H\textsubscript{9-n}} \\
\ce{C21H12+} &\longrightarrow \ce{C21H\textsubscript{12-n}+} \longrightarrow \ce{C8H5+} + \ce{C13H\textsubscript{7-n}} \\
\ce{C21H12+} &\longrightarrow \ce{C21H\textsubscript{12-n}+} \longrightarrow \ce{C9H3+} + \ce{C12H\textsubscript{9-n}} \\
\ce{C21H12+} &\longrightarrow \ce{C21H\textsubscript{12-n}+} \longrightarrow \ce{C10H5+} + \ce{C11H\textsubscript{7-n}} \\
\ce{C21H12+} &\longrightarrow \ce{C21H\textsubscript{12-n}+} \longrightarrow \ce{C11+} + \ce{C10H\textsubscript{12-n}} \\
\ce{C17H11+} &\longrightarrow \ce{C21H\textsubscript{12-n}+} \longrightarrow \ce{C10+} + \ce{C7H\textsubscript{1-n}}
\end{align*}

Since the mass visibility is more focused on the smaller cations it is clearly observed that \ce{C7H3+} and \ce{C8H5+} are the most favourable fragments of \ce{C21H12+}. It is interesting to note that a laser pulse energy of 2.581 mJ/pulse is sufficient to fragment the \ce{C21H12+} to produce \ce{C7H3+} and \ce{C8H5+} just after 2 laser pulses. The other cationic fragments shown in Figure \ref{subfig:1200_sum_BD} are in trace amounts; and most of them are produced at the beginning of the irradiation and keeps increasing till pulse 30.

\subsubsection{Theoretical results for the fragmentation of Sumanene cation}

Figure \ref{Sum_PES} shows the PES of the sumanene cation \ce{C21H12+}. For the theoretical section of sumanene, the different fragmentation channels identified is discussed followed by the explanation in the context of the experimental results. Because of structure and symmetry of the molecule (C$_{3v}$), there are only two type of H atoms considered in the first fragmentation: H atoms belonging to pentagons (blue circles in Figure \ref{Sum_PES}) and H atoms belonging to hexagons (red circles). The BDEs for these two H types are 3.95 and 4.96 eV, respectively, with the removal from the pentagon being favoured. The difference in energy reflects the different character of the \chemfig{C-H} bonds in the two cases, aromatic for the hexagons and aliphatic for the pentagons. A similar behaviour was reported both experimentally\cite{west2018unimolecular} and theoretically\cite{szczepanski2002preresonance} for the fluorene cation that also possesses a pentagon unit with 2H atoms. In those studies, the derived BDE for the H-loss from the pentagon was between 1.5 and 2 eV lower that one reported for sumanene here. H-loss can also happen after an isomerization reaction, in \textit{int1}, where the most likely H to be removed is the one attached to the tertiary carbon atom with a BDE of 2.07 eV. In fact, the H roaming to the tertiary carbon is energetically the most preferred route compared to the direct H loss that needs 3.95 eV.

The removal of the second H atom from the pentagon where the first H was removed requires 4.25 eV, while the removal of a second hydrogen from the party dehydrogenated hexagon requires slightly less, 4.00 eV (not in figure). If we consider that removing an H from any of the remaining pentagons requires also around 3.95 eV (see Figure \ref{Sum_PES}), this means that there is strong competition between the different H-loss channels and the peak corresponding to \ce{C21H10+} will be made of the isomer with both the H atoms abstracted from the sp$^3$ hybridized pentagon site. \ce{H2} can also be released from the aliphatic side with an energy barrier of 3.86 eV. As for corannulene, energetics are not enough to constrain if in the experiments we are witnessing \ce{H2} or 2H loss.

\subsubsection{Discussion - fragmentation of Sumanene cation}

\begin{figure}[h]
\centering
  \includegraphics[height=9.5cm]{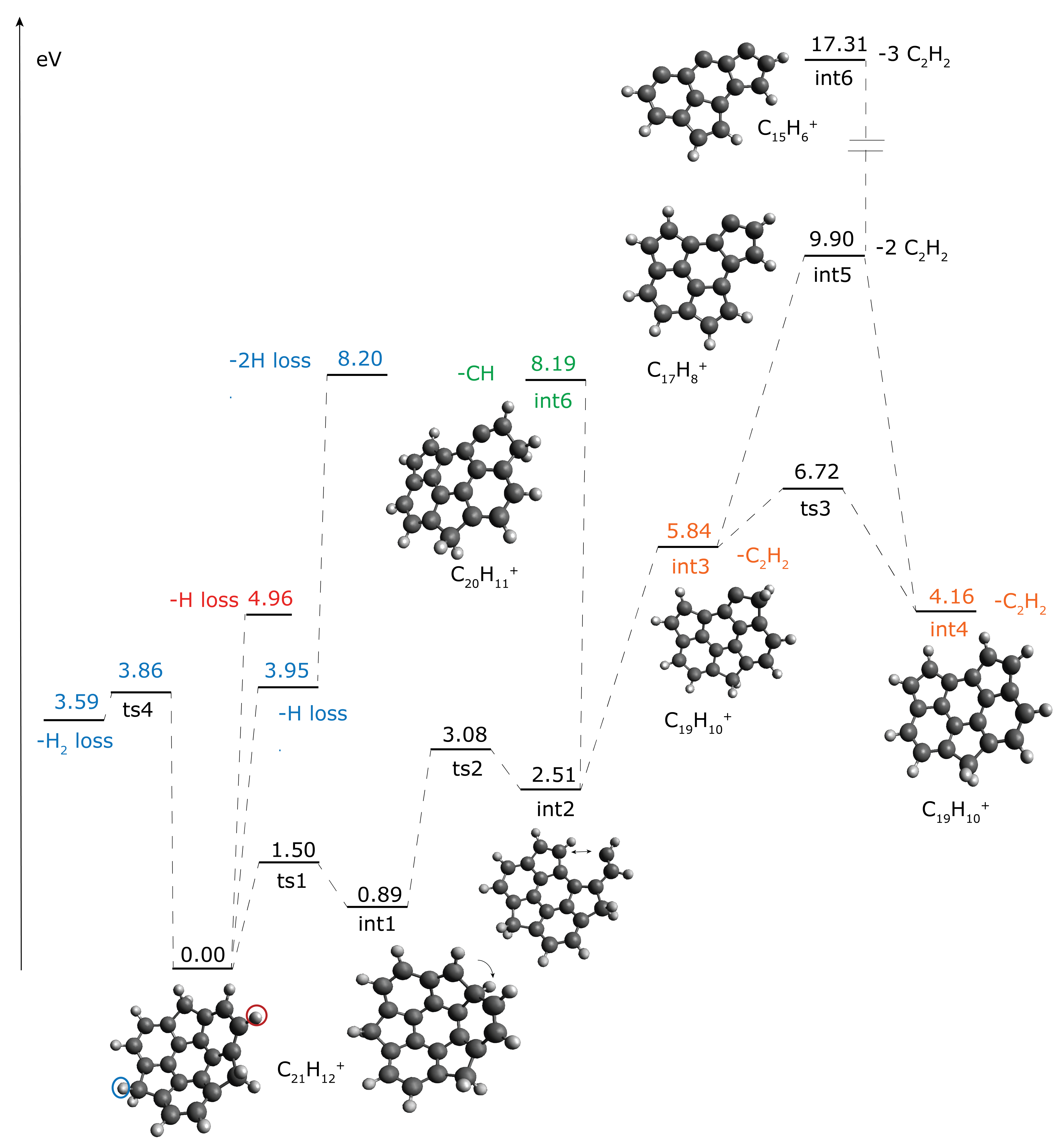}
  \caption{Potential Energy Surface (PES) for the \ce{C21H12+}, obtained at M06-2X/6-311++G(3df,2pd) level. Energies are  given relative to the energy of the sumanene cation. All the structures are in their lower multiplicity state (doublet), except for the structure resulting from the H-loss from the hexagon (triplet)and \textit{int6} (quadruplet). Dissociation channels are color-coded to help visualisation. In addition red highlight if the H atom is removed from a hexagon, blue  from a pentagon.}
  \label{Sum_PES}
\end{figure}

For the sumanene cations, C-loss channels pass through isomerization (Figure \ref{Sum_PES}), in particular H moving from the pentagon ring to a tertiary carbon (\textit{int1}, 0.89 eV) through a low barrier (\textit{ts1}, 1.5 eV). From there the breaking of a \chemfig{C-C} bond (\textit{ts2}, 2.19 eV with respect to \textit{int1}) leads to the formation of \textit{int2} (2.51 eV), where the hexagon opens up, creating a \ce{C2H2} unit that is attached to the nearby pentagon. Direct CH loss can happen from the \ce{C2H2} unit with a BDE of 5.68 eV (with respect to \textit{int2}), leaving behind a \ce{C20H11+} structure (\textit{int7}) where the remaining CH is inserted in the pentagon and forms a hexagon. Alternatively, only additional 3.33 eV are needed to release \ce{C2H2}; the resulting \\textit{int3} (\ce{C19H10+}) can isomerize to the more stable \textit{int4} (1.68 eV with respect to int3). Further \ce{C2H2} loss is expected to occur from \textit{int3} and \textit{int4} with a similar mechanism and almost the same BDE (5.74 eV for 2nd \ce{C2H2} versus 5.84 for the 1st ). \textit{Int5} (\ce{C17H8+}) is the first daughter of sumanene that is quasi-planar, and further loss of \ce{C2H2} from there requires almost 8 eV. The presence of several radical site in different rings is responsible of this high BDE value. We cannot exclude that additional structure rearrangement can take place, possibly lowering the BDE of the third \ce{C2H2} loss. This potential energy surface is in very good agreement with the experimental results where only \ce{C2H2} loss channels are observed.

\subsubsection{Comparison of the fragmentation pattern of buckybowls}

Sumanene fragmentation behaves much different from that of coronene and corannulene because of the presence of three sp$^3$ hybridized carbons associated with the pentagons in its peripheral structure, whereas each hexagon carbon in the periphery is bonded to only one hydrogen atom. The presence of pentagons affects the type of isomerization, leading sumanene to lose \ce{C2H2} rather than \ce{CCH2} as seen in the case of coronene and corannulene. Further, the presence of a pentagon seems to facilitate Stone-Wales like rearrangement (i.e. two neighbouring hexagons isomerizing to one pentagon and one 7-membered ring), in the structure of both corannulene and sumanene, which leads to more stable products and thus lower BDEs.
\vspace{0.5cm}

\textit{Fragmentation pattern of \ce{C20H10+} and \ce{C21H12+} compared to \ce{C24H12+}:} Coronene and corannulene have very similar periphery (hexagons with duo group) and similar isomerization mechanisms leading to C-loss channels but reaction barriers and intermediates are in general lower for corannulene. The experiments show different behaviour, such as mostly dehydrogenation for coronene and CH/\ce{C2H2} competition for corannulene. The different behaviour, when it comes to the competition at lower energy might be ascribed to the curvature of corannulene. H-roaming reaction facilitate ring-opening (see Figure \ref{Cor_PES1}), which in turns release the strain in the C–C bond of the corannulene structure. We expect daughter molecules of corannulene to be mostly flat as revealed by the predicted molecular structures. In principle, once the first fragmentations take place, the presence of a planar structure should reflect in the experiments which could be very similar to the initial fragments observed in coronene. It is interesting to note that corannulene lose \ce{CCH2} as shown in figure \ref{Cor_PES1}, whereas sumanene loses \ce{C2H2} as shown in figure \ref{Sum_PES}, though both are fragmented from a peripheral hexagon ring. This demonstrates that H-roaming plays a crucial role in the fragmentation process.

\textit{Similarities and differences between fragmentation of corannulene and sumanene cations:} The initial H losses were similar for both corannulene and sumanene – more favourable even-numbered H losses, and losses up to six H atoms upon photo dissociation. Corannulene cation displayed a competition between CH and \ce{C2H2} losses, whereas sumanene cation displayed only \ce{C2H2} losses. The most intense low mass cations produced by photodissociation of corannulene and sumanene were mostly similar (\ce{C7H3+}, \ce{C8H5+}, \ce{C9H3+}, \ce{C10H5+}). \ce{C11+} was observed in moderate intensity in these experiments.

The experimental and theoretical results demonstrate that peripheral H loss is more important than curvature in the photo-fragmentation process. It is evident that the number of H loss that eventually creates two or more radical sites is crucial to determine the H roaming mechanisms. The H roaming process in corannulene and sumanene is one of the main reasons why \ce{CCH2} loss is observed in corannulene, whereas \ce{C2H2} loss is observed in sumanene.

\subsection{Corannulene dication (\ce{C20H10++})}

\begin{figure*}[h]
    \centering
    \begin{subfigure}{9.5cm}
        \centering
        \includegraphics[width=\linewidth]{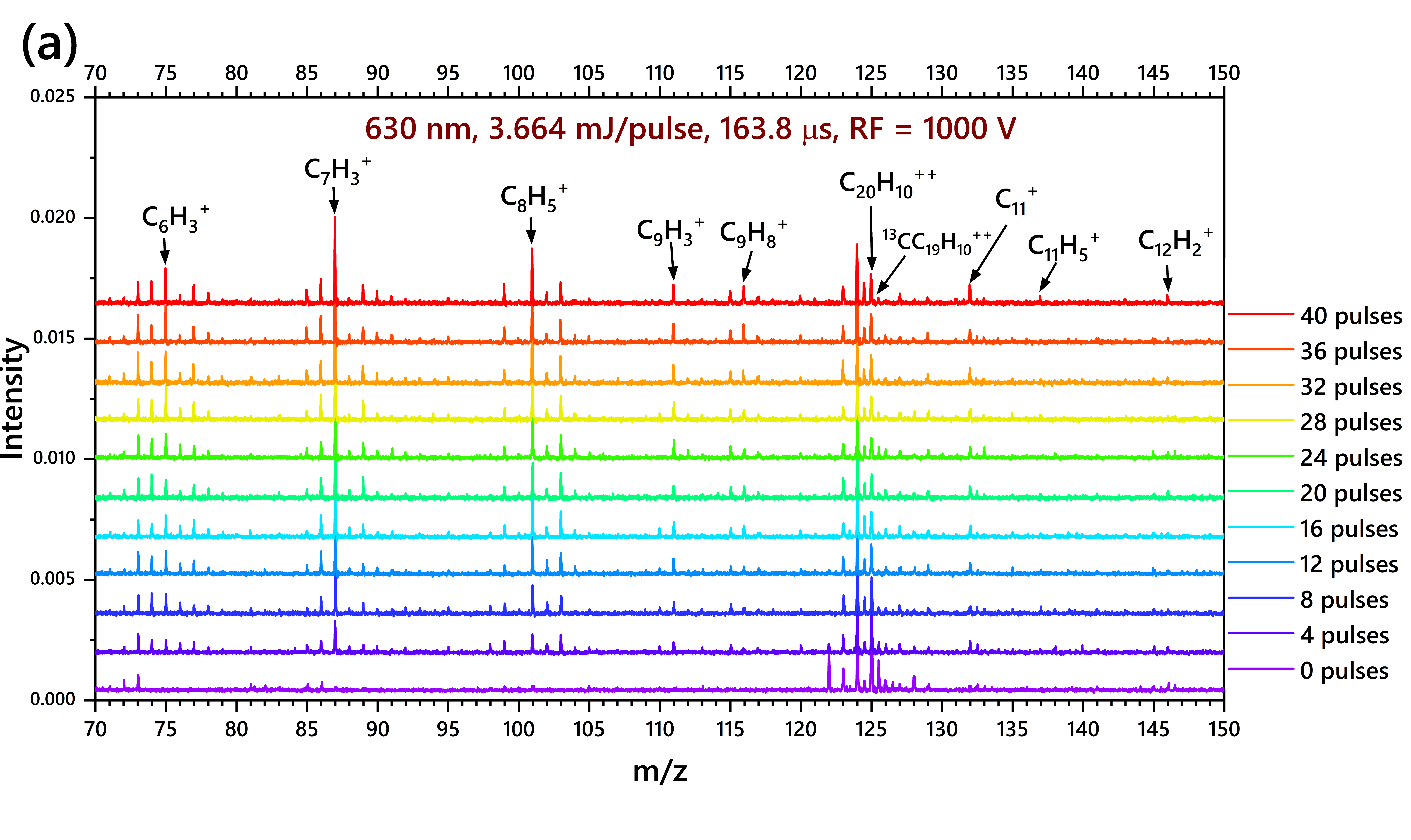}
        \phantomsubcaption
        \label{subfig:1000_cor++_MS}
    \end{subfigure}
   \hspace{\columnsep} 
    \begin{subfigure}{7.5cm}
        \centering
        \includegraphics[width=\linewidth]{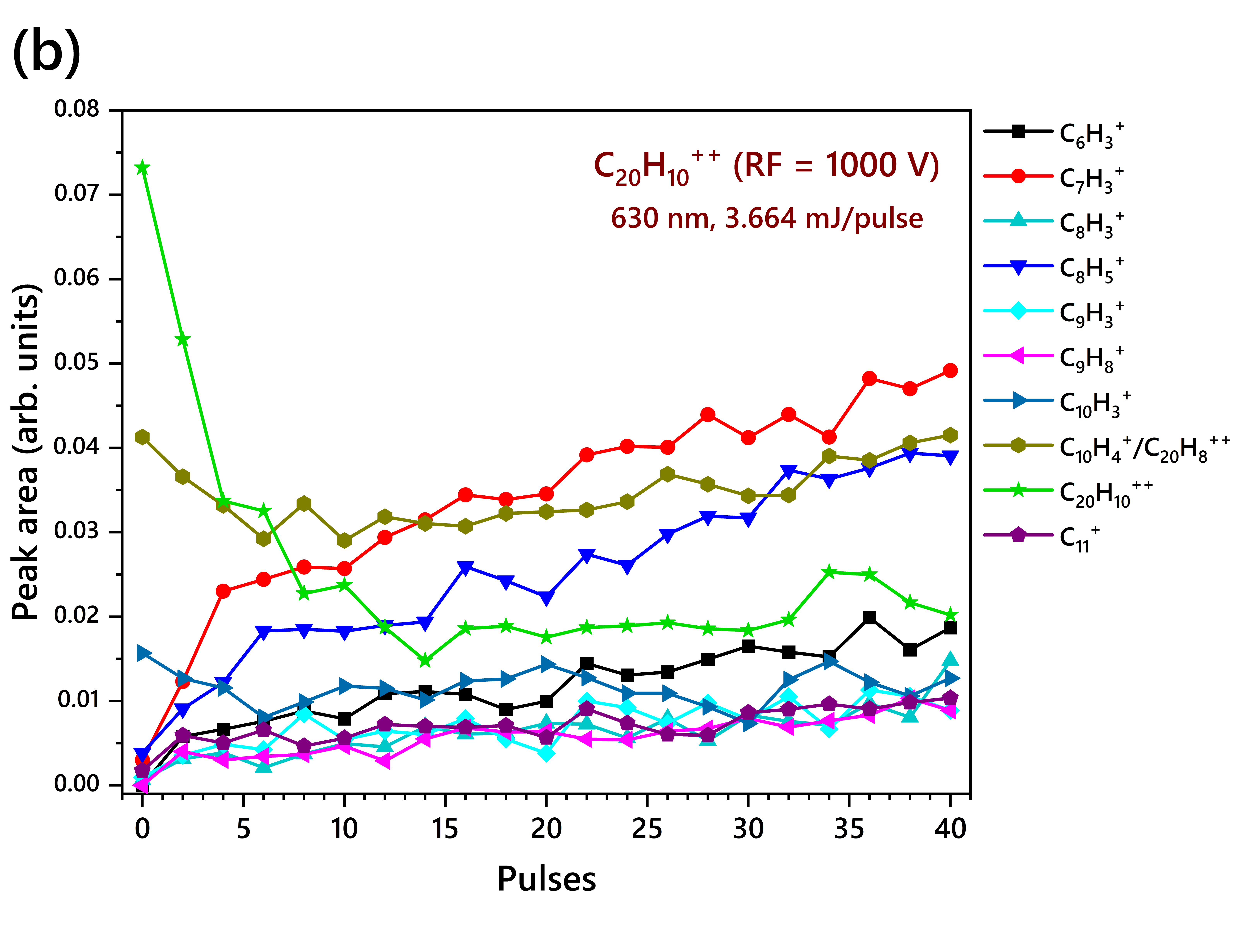}
        \phantomsubcaption
        \label{subfig:1000_cor++_BD}
    \end{subfigure}
   
    \caption{(a) TOF mass spectra for corannulene radical dications (\ce{C20H10++}) irradiated with 0, 2, 4,… and 40 laser pulses with a pulse energy of 3.664 mJ/pulse with RF = 1000 V i.e. spectra displaying the low-mass cationic fragments with prominent features at m/z = 150-70 range; (b) Normalized mass peak areas corresponding to the most prevalent hydrocarbon clusters formed in the fragmentation of the corannulene dication (\ce{C20H10++}).}
    \label{fig:1000_cor++}
\end{figure*}

\textit{Photodissociation of \ce{C20H10++} with 630 nm:} The electron energy when increased to 90 eV and emission of 0.65 mA produced intense peak of the corannulene dication, \ce{C20H10++} (m/z = 125). The SWIFT pulses were used to isolate the dication peaks and this was further subject to laser irradiation with red light from the dye laser at 630 nm, which provides the maximum energy of 3.664 mJ/pulse at a Q-switch delay of 163.8 µs which provides the optimum laser power for the fragmentation of dications (Figure \ref{subfig:1000_cor++_MS}). The photo-fragments produced after irradiation were all found to be cations and not dications. This means that the \ce{C20H10++} tends to fragment into two different cations. The parent peak \ce{(C20H10++)} at 125 m/z could possibly have contribution from \ce{C10H5+}, as the mass cannot be resolved to distinguish the two cations. Similarly, \ce{C20H6++} peak could have contributions from \ce{C10H3+}.

The major products observed in the experiments \ce{C6H3+}, \ce{C7H3+}, \ce{C8H5+}, \ce{C9H3+}, \ce{C9H8+} and \ce{C11+}. It was not possible to observe any cations beyond 150 m/z because of the low RF value of 1000 V. The very high abundance of \ce{C6H3+} and \ce{C7H3+} suggests that \ce{C20H10++} is capable of fragmenting into more than two components, following these mechanisms:
\begin{align*}
\ce{C20H10++} &\longrightarrow \ce{C7H3+} + \ce{C8H5+} + \ce{C5H2} \\
\ce{C20H10++} &\longrightarrow \ce{C6H3+} + \ce{C8H5+} + \ce{C6H2} \\
\ce{C20H10++} &\longrightarrow \ce{C6H3+} + \ce{C7H3+} + \ce{C7H4} \\
\ce{C20H10++} &\longrightarrow \ce{C9H3+} + \ce{C8H5+} + \ce{C3H2} \\
\ce{C20H10++} &\longrightarrow \ce{C9H3+} + \ce{C7H3+} + \ce{C4H4} \\
\ce{C20H10++} &\longrightarrow \ce{C9H3+} + \ce{C6H3+} + \ce{C5H4} \\
\ce{C20H10++} -\ce{H2} &\longrightarrow \ce{C20H8++} \longrightarrow \ce{C9H8+} + \ce{C11+}    
\end{align*}

The breakdown equations predicted above are based on the data obtained in the mass spectrum and the breakdown diagram. The breakdown of this dication into the low mas monocations are expected to involve several intermediate steps. The equations provided do not include any intermediate or transition states, as molecular dynamics simulations alone can elucidate them. Moreover, the corannulene dications were not found to have any observable dicationic fragment (though dicationic fragments could be present in trace amounts) based on the m/z analysis. Which means that the majority of the fragments are more likely to be cationic species. The structure identification, (except the structure of \ce{C7H3+} which is already known \cite{chakraborty2014spectroscopic}) and the elucidation of the breakdown sequence requires sophisticated molecular dynamics calculations, which is a future scope of this research.

\subsection{Sumanene dication (\ce{C21H12++})}

\begin{figure*}[h]
    \centering
    \begin{subfigure}{9.5cm}
        \centering
        \includegraphics[width=\linewidth]{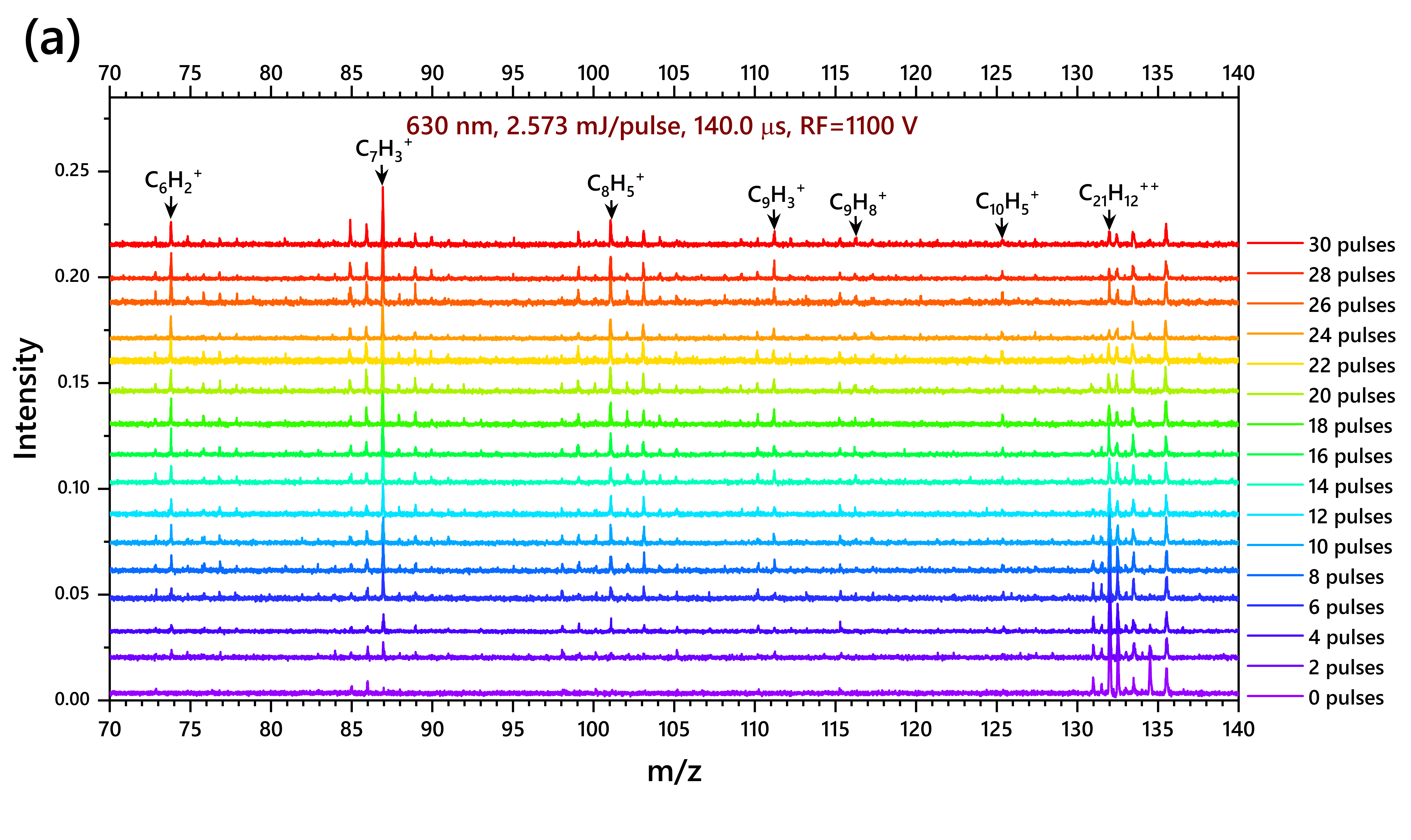}
        \phantomsubcaption
        \label{subfig:1000_sum++_MS}
    \end{subfigure}
   \hspace{\columnsep} 
    \begin{subfigure}{7.5cm}
        \centering
        \includegraphics[width=\linewidth]{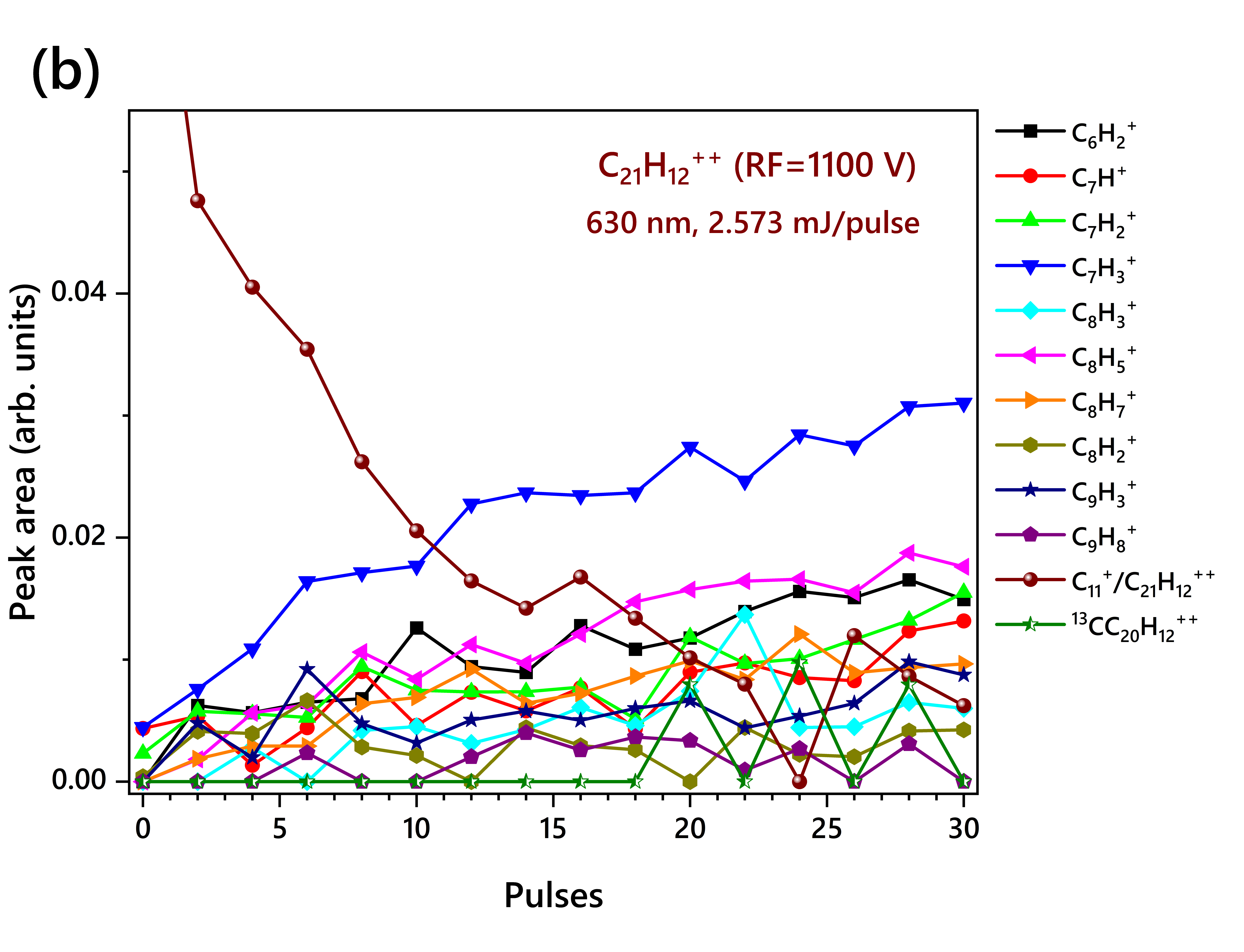}
        \phantomsubcaption
        \label{subfig:1000_sum++_BD}
    \end{subfigure}
   
    \caption{(a) TOF mass spectra for sumanene radical dications irradiated with 0, 2, 4,… and 30 laser pulses with a pulse energy of 2.573 mJ/pulse with RF = 1100 V i.e. spectra displaying the low-mass cationic fragments with prominent features at m/z = 140-70 range; (b) Normalized mass peak areas corresponding to the most prevalent hydrocarbon clusters formed in the fragmentation of the sumanene dication (\ce{C21H12++}).}
    \label{fig:1000_sum++}
\end{figure*}

\textit{Photodissociation of \ce{C21H12++} with 630 nm:} The electron energy when increased to 80 eV and emission of 0.65 mA produced intense peak of the sumanene dication, \ce{C21H12++} (m/z = 132). The SWIFT pulses were used to isolate the dication peaks and was further subject to laser irradiation at 630 nm, which provides an energy of 2.573 mJ/pulse at a Q-switch delay of 140.0 µs (Figure \ref{subfig:1000_sum++_MS}). The photo-fragments produced after irradiation were only cations just like the case of corannulene dication. \ce{C6H2+}, \ce{C7H3+}, \ce{C8H5+}, \ce{C9H3+} are the most abundant products of fragmentation of \ce{C21H12++}. Again, the mass beyond 140 m/z could not be observed because of the constrain of the RF value (1100 V). The very high abundance of \ce{C6H2+} and \ce{C7H3+} suggests that \ce{C21H12++} is also capable of fragmenting into more than two components as follows:
\begin{align*}
\ce{C21H12++} &\longrightarrow \ce{C7H3+} + \ce{C8H5+} + \ce{C6H4} \\
\ce{C21H12++} &\longrightarrow \ce{C6H3+} + \ce{C8H5+} + \ce{C7H4} \\
\ce{C21H12++} &\longrightarrow \ce{C6H3+} + \ce{C7H3+} + \ce{C8H6} \\
\ce{C21H12++} &\longrightarrow \ce{C9H3+} + \ce{C8H5+} + \ce{C4H4} \\
\ce{C21H12++} &\longrightarrow \ce{C9H3+} + \ce{C7H3+} + \ce{C5H6} \\
\ce{C21H12++} &\longrightarrow \ce{C9H3+} + \ce{C6H3+} + \ce{C6H6}
\end{align*}

The sumanene dications were not found to have any observable dicationic fragment. So, the majority of the fragments are more likely to be cationic species just like the case of corannulene. Figure \ref{subfig:1000_sum++_BD} shows the breakdown diagram of the cations produced upon irradiation the \ce{C21H12++}. The parent dication has the same mass as that of \ce{C11+} (m/z = 132) which cannot be resolved with the mass resolution of this TOF-MS, but is expected to be produced as one of the significant products in almost all the PAHs experimented with the i-PoP system with C<14. Hence the trace of the peak area of the \ce{C21H12++} does not drop to zero but keeps growing after pulse 24.

\section{Astrophysical Implications}

The clue to the understanding of the relationship between \ce{C60} and PAHs in space could be established by studying the molecules that could be intermediate steps leading  to the formation of \ce{C60} from PAHs.\cite{zhen2014laboratory} This is where the Buckybowls gain importance in astrochemistry. Theoretical calculations and molecular dynamics simulations have shown that a fully benzenoid planar PAH can undergo dehydrogenation and possess one or more radical sites, which in turn pave the way for isomerization under intense radiation fields to form a pentagon rings through Stone-Wales like defects.\cite{stein1985high,berne2012formation} This effect was also experimentally observed in one of the previous work in our laboratory where photodissociation of \ce{C66H28} produced the fullerene, \ce{C60}. Such PAHs with pentagons are also suggested to be more stable at interstellar conditions.\cite{narahariasastry1993synthetic} Alternatively, pentagons could be formed in space through CO loss from methoxy-functionalized PAHs,\cite{zhen2016photo} produced by photolysis of PAHs trapped in interstellar ices. In view of this, molecular dynamics of pentagon containing PAHs, such as corannulene and sumanene, becomes imperative to link the bridge between the \ce{C60} and PAHs. Moreover, these are medium-sized PAHs with C $\geq$ 20 that are symmetric (C$_{5v}$ and C$_{3v}$ for Corannulene and Sumanene respectively) and stable. Moreover, these are close in size and structure to the fully benzenoid PAH, coronene (\ce{C24H12}) hat is commonly studied as a typical PAH representative in molecular astrophysics.

The AIBs are particularly bright in so-called Photodissociation region (PDRs) in space. PDRs are regions in space where far-UV photons with energies between 6 and 13.6 eV penetrate molecular clouds, photo-dissociating molecules and ionizing atoms such as C and heating the gas to temperatures of ~500 K. Spectroscopic observations have revealed that the characteristics of the emitting interstellar PAH family changes with location in the PDR. This has been attributed to the photochemistry and photo-isomerization produced by the penetrating far-UV photons,\cite{berne2012formation,joblin1996spatial,chown2024pdrs4all} leading to the loss of functionalized groups such as \ce{CH3} and of aromatic H, and the isomerization to \ce{C60} and other fullerenes and cages. The surfaces of these PDRs also contain abundant small hydrocarbon radicals, and it has been suggested that these are the photochemically daughter products of PAHs exposed to the strong far-UV radiation field.\cite{pety2005pahs}  

This study on the photofragmentation of corannulene and sumanene demonstrates that the buckybowl cations in the photodissociation region can undergo fragmentation to produce a variety of high and low mass fragment cations, which may contribute to the rich diversity of the interstellar molecular inventory. These experiments reveals the top-down approach where large PAHs breakdown to form smaller hydrocarbons or carbon skeletons. Interestingly, the fragment cations like \ce{C10+}, \ce{C11+}, etc., produced in these experiments are expected to have ring structures and are more stable than the fragments with linear or branched structures. It is to be noted that the carbon clusters like \ce{C10+}, \ce{C11+}, \ce{C12+}, \ce{C13+}, \ce{C14+}, etc., were produced predominantly for PAHs like coronene, dibenzopyrene, etc. But the buckybowl cations produce only \ce{C10+} and \ce{C11+}, and only in trace amounts. This indicates the important influence of planarity on the photo-fragmentation of a PAH. The number of radical sites and the ring opening mechanism in the carbon skeleton are crucial in deciding the formation of such ring-structured carbon cations. The low mass cation \ce{C7H3+} and \ce{C8H5+} were produces in high abundance for both cationic and dicationic buckybowls. These small cations could be present in high abundance in the ISM. Given the intense radiation fields in the ISM, the cationic fragments \ce{C7H3+}, \ce{C8H5+}, \ce{C9H3+}, \ce{C10H5+}, \ce{C10+}, \ce{C11+}, etc., could potentially be the proxies to hunt for PAHs including the buckybowls (and other mid- to large-sized PAHs) in space. 

The absorption spectrum of the interstellar medium contains a set of ~500 absorption bands, the so-called diffuse interstellar bands (DIBs), in the near-UV to far-red range. These DIBs are generally attributed to molecular electronic absorptions.\cite{tielens2012diffuse,cox2013road} Indeed, five of these bands have been identified with the fullerene, \ce{C60+}.\cite{campbell2015laboratory} The photochemical products of the PAHs in the experiments reported here could also contribute to the DIBs. Further laboratory studies are warranted to characterize the rotational, vibrational, and electronic spectroscopic properties of these species and to enable a search for them in space.

Finally, we note that, though single photon excitations dominate in space, multi photon processes have a non-negligible probability to occur as some hundred million absorption events will happen over the lifetime of an interstellar PAH.\cite{tielens2005physics} As a consequence of these process, the PAHs either emit IR fluorescence, fragment, or isomerize. This rich inventory of pentagon containing buckybowl cations, dications and their photo-fragmented low mass cations are important candidates to search for with the James Webb Space Telescope. 

\section{Conclusion}

This work demonstrates that a diverse variety of fragment cations can be produced by photo dissociation of a PAH cation in an ion trap. The TOF mass spectrum provides information about the various high and low mass cations that are produced by irradiation of corannulene and sumanene cations with a red light from a dye laser. Corannulene was found to lose up to six H atoms initially, where even H losses were found to be more favourable than the odd ones. This is followed by a competition between CH and \ce{CCH2} losses. The competition between the CH and \ce{CCH2} loses is evident in the comparable energy difference of 0.09 eV, as estimated from the calculated potential energy diagram (7.45 eV and 7.54 eV respectively). This means that the energy to lose CH and \ce{CCH2} can almost be considered equal considering the errors in the predicted energies. 

Sumanene also lost up to six H atom which was clearly followed by only \ce{C2H2} losses. The potential energy diagram displays an energy difference of 2.71 eV between the \ce{C2H2} and CH losses (5.84 and 8.19 eV respectively for the \ce{C2H2} and CH losses) which is the reason why only \ce{C2H2} losses are observed in the case of sumanene. Interestingly, the most intense low mass cations produced by photodissociation of corannulene and sumanene cations were mostly similar (\ce{C7H3+}, \ce{C8H5+}, \ce{C9H3+}, \ce{C10H5+}). The photodissociation of corannulene and sumanene dications were also found to produce intense \ce{C7H3+}, \ce{C8H5+}, \ce{C9H3+}, \ce{C10H5+} cation, which reveals that the fragmentation process is irrespective of the charge state of a PAH. The high abundance of these small hydrocarbons in the experiments described in this paper is interesting and calls for similar experiments, down to low mass fragments, also for other PAH cations to understand if these species can indeed be the end point of top-down interstellar chemistry of PAHs. In addition, experimental and theoretical characterisation of the  spectroscopy of these species (eg. \ce{C7H3+} \cite{chakraborty2014spectroscopic}) is fundamental to search for them in the ISM.

These results provide an insight into the evolution of PAHs with pentagon(s) in space. These experiments determine the vulnerability of pentagon-bearing PAHs under intense radiations in space condition and will aid to identify possible fragments of PAHs with pentagon rings that could be present in the ISM. The future scope of this work is to provide the IR characteristics a of such species that can be searched for in the JWST data.

\section*{Author Contributions}

PS – Conceptualization, Data curation, Formal analysis, Funding acquisition, Investigation, Methodology, Validation, Visualization, Writing – original draft, Writing – review and editing. \\
AC – Conceptualization, Formal analysis, Investigation, Methodology, Validation, Visualization, Writing – original draft, Writing – review and editing. \\
JK – Investigation, Methodology, Validation, Software, Writing – review and editing. \\
HL – Supervision, Investigation, Visualization, Validation. \\
AT – Supervision, Conceptualization, Funding acquisition, Investigation, Validation, Writing – review and editing.

\section*{Conflicts of interest}
There are no conflicts to declare.

\section*{Data availability statement}
The data supporting this article have been included as part of the Supplementary Information. See DOI: 10.1039/cXCP00000x/.

\section*{Acknowledgements}
P.S. acknowledges the European Union and Horizon 2020
Postdoctoral funding awarded under the Marie Skłodowska-Curie action (grant number 101062984) H.L. acknowledges the Netherlands Research School for Astronomy (Nederlandse Onderzoekschool Voor Astronomie, NOVA) and the Netherlands Organisation for Scientific Research (Nederlandse Organisatie voor Wetenschappelijk Onderzoek, NWO) A.G.G.M.T. acknowledges the NWO for a Spinoza Prize (Spinozapremie). We thank SURF (\url{www.surf.nl}) for the support in using the National Supercomputer Snellius.



\balance


\bibliography{rsc} 
\bibliographystyle{rsc} 

\end{document}